\documentclass{aa}  

\usepackage{graphicx}
\usepackage{txfonts}
\usepackage[colorlinks=true, citecolor=blue]{hyperref}
\usepackage{booktabs}

\newcommand{\es}{erg s$^{-1}$}

\newcommand{\kms}{km s$^{-1}$}

\newcommand{\msun}{$M_{\odot}$}
\newcommand{\mstar}{$M_{\star}$}
\newcommand{\mgas}{$M_{\rm gas}$}
\newcommand{\fgas}{$f_{\rm gas}$}
\newcommand{\mdust}{$M_{\rm dust}$}
\newcommand{\kkmspc}{K~km~s$^{-1}$~pc$^2$}

\newcommand{\jykms}{Jy~km~s$^{-1}$}
\newcommand{\ci}{[C\,{\footnotesize I}]}

\newcommand{\co}{CO}
\newcommand{\cione}{[C\,{\footnotesize I}]$(^3P_1\,-\,^{3}P_0)$}
\newcommand{\citwo}{[C\,{\footnotesize I}]$(^3P_2\,-\, ^{3}P_1)$}

\newcommand{\cofive}{CO\,$(5-4)$}
\newcommand{\coseven}{CO\,$(7-6)$}
\newcommand{\cofour}{CO\,$(4-3)$}

\newcommand{\cothree}{CO\,$(3-2)$}
\newcommand{\cotwo}{CO\,$(2-1)$}

\newcommand{\lprimecione}{$L'_{\mathrm{[C\,\scriptscriptstyle{I}\scriptstyle{]}}^3P_1\,-\, ^3P_0}$}

\newcommand{\lprimecitwo}{$L'_{\mathrm{[C\,\scriptscriptstyle{I}\scriptstyle{]}}^3P_2\,-\, ^3P_1}$}
\newcommand{\lprimecotwo}{$L'_{\rm CO(2-1)}$}

\newcommand{\lprimecofour}{$L'_{\rm CO(4-3)}$}
\newcommand{\lprimecofive}{$L'_{\rm CO(5-4)}$}
\newcommand{\lprimecoseven}{$L'_{\rm CO(7-6)}$}

\newcommand{\lir}{$L_{\rm IR}$}
\newcommand{\lirsfr}{$L_{\rm IR,\,SFR}$}
\newcommand{\liragn}{$L_{\rm IR,\,AGN}$}

\newcommand{\lsun}{$L_{\odot}$}

\newcommand{\umean}{$\langle U \rangle$}

\newcommand{\fagn}{$f_{\rm AGN}$}
\newcommand{\lx}{$L_{2-10\,\mathrm{keV}}$}
\newcommand{\ratio}{$R_{52}$}

\begin{document}

   \title{The effect of active galactic nuclei on the cold
     interstellar medium in distant star-forming galaxies}

   \author{F. Valentino\inst{1,2},
          E. Daddi\inst{3},
          A. Puglisi \inst{4},
          G. E. Magdis \inst{1,5,6},
          V. Kokorev \inst{1,2},
          D. Liu \inst{7},
          S. C. Madden \inst{3},
          C. G\'{o}mez-Guijarro \inst{3},
          M.-Y. Lee \inst{8},
          I. Cortzen \inst{9},
          C. Circosta \inst{10},
          I. Delvecchio \inst{11}
          J. R. Mullaney \inst{12},
          Y. Gao \inst{13,14},
          R. Gobat \inst{15},
          M. Aravena \inst{16},
          S. Jin \inst{17,18},
          S. Fujimoto \inst{1,2}, 
          J. D. Silverman \inst{19,20} \and
          H. Dannerbauer \inst{13,14}
     }
     \institute{Cosmic Dawn Center (DAWN), Denmark\\
          \email{francesco.valentino@nbi.ku.dk}
          \and Niels Bohr Institute, University of
          Copenhagen, Jagtvej 128 DK-2200 Copenhagen N, Denmark
          \and AIM, CEA, CNRS, Universit\'e Paris-Saclay, Universit\'e
          Paris Diderot, Sorbonne Paris Cit\'e, F-91191
          Gif-sur-Yvette, France
          \and Center for Extragalactic Astronomy, Durham University,
          South Road, Durham DH13LE, United Kingdom
          \and DTU-Space, Technical University of Denmark, Elektrovej
          327, DK-2800 Kgs. Lyngby, Denmark
          \and Institute for Astronomy, Astrophysics, Space
          Applications and Remote Sensing, National Observatory of
          Athens, GR-15236 Athens, Greece
          \and Max Planck Institute for Extraterrestrial Physics
          (MPE), Giessenbachstr. 1, 85748 Garching, Germany 
          \and Korea Astronomy and Space Science Institute, 776
          Daedeokdae-ro, 34055 Daejeon, Republic of Korea
          \and Institut de Radioastronomie Millim\'{e}trique (IRAM), 300 rue de la
          Piscine, 38400 Saint-Martin-d’H\`{e}res, France
          \and Department of Physics \& Astronomy, University College
          London, Gower Street, London, WC1E 6BT, UK
          \and INAF -- Osservatorio Astronomico di Brera, via Brera 28,
          I-20121, Milano, Italy
          \and Department of Physics and Astronomy, The University
          of Sheffield, Hounsfield Road, Sheffield, S3 7RH, UK 
          \and Department of Astronomy, Xiamen University, Xiamen,
          Fujian 361005, People’s Republic of China
          \and  Purple Mountain Observatory \& Key Laboratory for Radio
          Astronomy, Chinese Academy of Sciences, 10 Yuanhua Road,
          Nanjing 210033, People’s Republic of China 
          \and Instituto de F\'{i}sica, Pontificia Universidad Cat\'{o}lica de
          Valpara\'{i}so, Casilla 4059, Valpara\'{i}so, Chile 
          \and N\'{u}cleo de Astronom\'{i}a, Facultad de Ingenier\'{i}a y
          Ciencias, Universidad Diego Portales, Av. Ej\'{e}rcito 441,
          Santiago, Chile
          \and Instituto de Astrof\'{i}sica de Canarias (IAC), E-38205
          La Laguna, Tenerife, Spain
          \and Universidad de La Laguna, Dpto. Astrof\'{i}sica,
          E-38206 La Laguna, Tenerife, Spain 
          \and Kavli Institute for the Physics and Mathematics of the
          Universe (Kavli-IPMU, WPI), The University of Tokyo
          Institutes for Advanced Study, The University of Tokyo,
          5-1-5 Kashiwanoha, Kashiwa, Chiba 277-8583, Japan
          \and Department of Astronomy, School of Science, The
          University of Tokyo, 7-3-1 Hongo, Bunkyo-ku, Tokyo 113-0033,
          Japan
          }
  \authorrunning{Valentino et al.}
  \titlerunning{}
   \date{Received --; accepted --}
 
  \abstract
  {In the framework of a systematic study with the ALMA
    interferometer of IR-selected main-sequence and starburst galaxies at $z\sim1-1.7$ at typical $\sim1"$ resolution, we
    report on the effects of mid-IR- and X-ray-detected active galactic
    nuclei (AGN) on the reservoirs and excitation of molecular gas
    in a sample of 55 objects. We find
    widespread detectable nuclear activity in $\sim 30$\% of the
    sample. The
    presence of dusty tori influences the IR
    spectral energy distribution of galaxies, as highlighted by the
    strong correlation among the AGN contribution to the total IR
    luminosity budget ($f_{\rm AGN}=L_{\rm IR,\,AGN}/L_{\rm IR}$),
    its hard X-ray emission, and the Rayleigh-Jeans to mid-IR ($S_{\rm 1.2\,mm}$/$S_{\rm
      24\mu m}$) observed color, with evident consequences on
    the ensuing empirical star formation rate estimates. Nevertheless, we find only
    marginal effects of the presence and strength of AGN on the carbon
    monoxide CO ($J=2,4,5,7$) or neutral carbon
    (\cione, \citwo) line luminosities and on the derived molecular gas
    excitation as gauged by line ratios and the full spectral line
    energy distributions. The \ci\ and CO emission up to $J=5,7$ thus
    primarily traces the properties of the host in typical IR luminous
    galaxies.
    However, our analysis highlights the
    existence of a large variety of line luminosities and ratios despite the
    homogeneous selection. In particular, we find a sparse group of AGN-dominated
    sources with the highest $L_{\rm IR,\,AGN}/L_{\rm
      IR,\,SFR}$ ratios, $\gtrsim3$,  that are more luminous in \cofive\ than
    what is predicted by the \lprimecofive-\lirsfr\ relation, which might
    be the result of the nuclear activity.
    For the general population, our findings translate into AGN having
    minimal effects on
    quantities such as gas and dust fractions and star formation
    efficiencies. If anything, we find hints of a marginal tendency
    of AGN hosts to be compact at far-IR wavelengths and to display $1.8$ times
    larger
    dust optical depths. In general, this is
    consistent with a marginal impact of the
    nuclear activity on the gas reservoirs and star formation in
    average star-forming AGN hosts with $L_{\rm IR}>5\times10^{11}$ \lsun, typically underrepresented in surveys
    of quasars and submillimeter galaxies.}

   \maketitle
%

\section{Introduction}
\label{sec:introduction}
The role of super-massive black holes (SMBHs) in the evolution of their host
galaxies is a topic of major debate in extragalactic
astrophysics (see \citealt{harrison_2017} for a recent review). The empirical correlations among several properties of
SMBHs and of their hosts suggest a tightly intertwined evolution, which has
its physical roots in the mechanisms that trigger their growth. From a theoretical
standpoint, the phases of active growth of SMBHs -- shining as active
galactic nuclei (AGN) -- are regarded as primary regulators of the hosts'
growth and a viable mechanism to completely quench the star formation
\citep[e.g.,][]{croton_2006, hopkins_2006}. This might happen via the expulsion
of a significant portion of the cold gas reservoirs via powerful winds
or by heating the interstellar medium (ISM) and the intergalactic medium,
preventing further gas collapse and accretion (``negative feedback''). Interestingly, the
same AGN-driven outflows might compress the surrounding gas and
stimulate, rather than suppress, the star formation in the host
\citep[``positive feedback,'' e.g.,][]{silk_2013}. It is, thus,
evident that the study of star formation in galaxies, cold gas
reservoirs, and their physical conditions has a central role in this context. However,
establishing a causality nexus between the AGN activity and the galaxy
growth or quenching has
turned out to be a significant challenge owing to the widely different
time and spatial scales involved in the development
of SMBHs and their hosts, population
selection biases, the sparsity of the available samples, and the truly
multivariate nature of this problem, as different processes concur to
explain common sets of observables. 

As an example, observational signatures of high-velocity and multiphase
outflows launched by AGN have been reported in the local and distant
Universe \citep[e.g.,][]{cicone_2014, cicone_2018,
  forster-schreiber_2014, feruglio_2015, harrison_2015, jarvis_2019, veilleux_2020}, and so have pockets of warm and dense gas in AGN
hosts \citep[e.g.,][]{weiss_2007, vanderwerf_2010, mashian_2015,
  carniani_2019}. Nevertheless, there is no clear evidence of 
strong feedback on the cold gaseous and dusty ISM in sizable
samples of local galaxies, with AGN hosts showing similar global properties
to purely star-forming galaxies \citep{saintonge_2012, shangguan_2020,
  jarvis_2020, yesuf_2020}. 
At high redshifts the situation is more controversial. 
Past studies particularly focused on the lowest accessible carbon
monoxide (CO) transitions and long-wavelength dust emission in order
to constrain the total
molecular gas mass across different galaxy types. A great effort
has been undertaken to characterize the epoch of 
maximal growth of galaxies via star formation and subsequent quenching at
$z\sim1-3$, when the cosmic SMBH accretion rate peaked \citep{delvecchio_2020}.
This resulted in claims of null effects of AGN on the star formation rates
\citep[SFRs;][]{stanley_2017, schulze_2019} and the ISM properties of the host
\citep{sharon_2016, kirkpatrick_2019} or, at the opposite end, decreased gas fractions and
shortened depletion timescales in the presence of AGN \citep{kakkad_2017, perna_2018, brusa_2018,
circosta_2021, bischetti_2021},
disfavoring and supporting the existence of some kind of AGN feedback,
respectively. 
\begin{figure*}
\includegraphics[width=\textwidth]{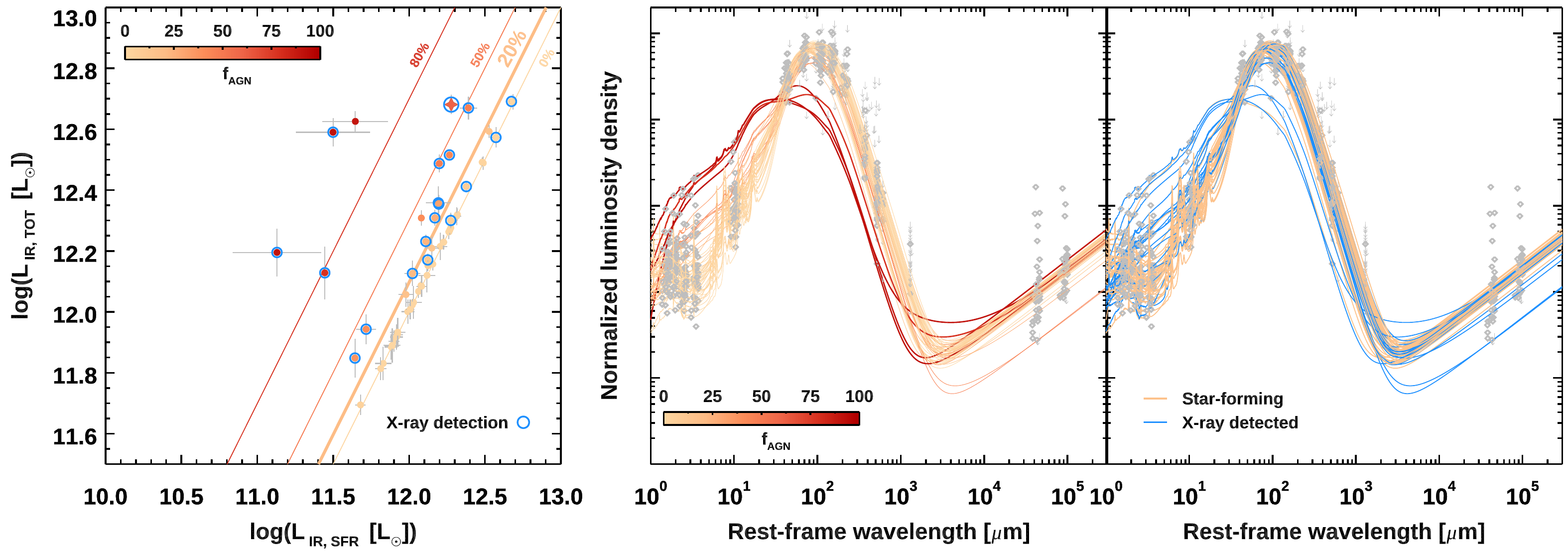}
\caption{{Star-formation and AGN contributions to the total
    IR luminosities $L_{\rm IR}(8-1000\,\mu \mathrm{m})$.}
  \textit{Left:} Our sample of galaxies at $z\sim1.2$, indicated by filled circles
  and color coded according
  to the AGN contribution to \lir\ ($f_{\rm AGN}=L_{\rm
    IR,\,AGN}/L_{\rm IR}$). Blue circles show hard X-ray
  emitters with $L_{2-10\,\mathrm{keV}}>10^{43}$ \es\ detected by
  \textit{Chandra}. The orange filled star marks the object from
  \cite{brusa_2018}, which is part of our parent sample. The colored lines show the limits for \fagn\ as
  labeled. \textit{Center, right:} Rest-frame far-IR SEDs normalized to
  $L_{\rm IR}(8-1000\,\mu \mathrm{m}) = 1 $ \lsun\ and color coded
  according to \fagn\ (central panel) and X-ray detection (right
  panel). Gray diamonds and arrows show detections and $3\sigma$ upper
  limits on the photometry, respectively. We note that the 
  emission at radio wavelengths is not included in the modeling.}
\label{fig:fig1}
\end{figure*}

If the inferred total amount of molecular gas is disputed,
even less is known about the gas excitation conditions in AGN hosts except for the brightest -- and most
biased -- quasars or nearby (Ultra-)Luminous InfraRed Galaxies \citep[(U)LIRGs,][]{rosenberg_2015}. A better
knowledge of the gas properties is highly desirable
as it would have an immediate impact on the disparate conclusions about gas
masses, fractions, and star formation efficiencies (SFEs) mentioned above,
considering that an excitation correction is very frequently required to scale
the observed mid- or high-$J$ CO emission to the ground
transition. Measurements of the CO spectral line energy
distribution (SLED) of well-known powerful quasars
(QSOs) at high redshift have revealed rather extreme luminosity ratios
\citep[e.g.,][]{weiss_2007, carilli_2013, bischetti_2021}
and detections of very high-$J$ transitions
\citep{gallerani_2014, carniani_2019}. However, it is not clear how
different these ISM conditions are compared with the ones of starbursting
submillimeter galaxies (SMGs; e.g., 
\citealt{bothwell_2013, spilker_2014}) generally observed at similar
redshifts or at less extreme and more common AGN luminosities
\citep{sharon_2016, kirkpatrick_2019}. Part of the confusion stems from the
existence of several mechanisms able to similarly excite CO and
produce rather flat SLEDs at high $J$, such as X-rays \citep{meijerink_2007,
  vanderwerf_2010}, mechanical heating \citep{rosenberg_2015}, shocks induced by mergers, radio jets, and supernova- or
AGN-driven outflows \citep{kamenetzky_2016, carniani_2019}, and
turbulence \citep{harrington_2021}. The picture is
complicated by the limited amount 
of information for statistically significant samples that encompass
both AGN hosts and control star-forming galaxies matched in
stellar mass, SFR, and other relevant properties. Systematic studies
of gas excitation conditions could thus help to shed light on the
coevolution of AGN and their hosts.

Here we attempt to characterize the AGN activity in the population of
main-sequence galaxies and strong starbursts at
$z\sim1-1.7$ that we surveyed in several CO and \ci\ transitions with the Atacama Large Millimeter
Array (ALMA; \citealt{puglisi_2019, puglisi_2021, valentino_2020c} -- V20a from
hereafter). Our goal is to gauge the effect of accreting SMBHs on the global
SFR, the mass, and the excitation conditions of the cold gas and dust in
a large sample of homogeneously IR-selected
galaxies. This results in a focus on the AGN activity
spread in normal galaxies that are representative of the bulk of the
star-forming population \citep{rodighiero_2011}, rather than on the
extreme SMGs and QSOs investigated in depth in previous works. For
this purpose, we take advantage of a simple and well-understood selection function, the large sample of a few
tens of galaxies observed under the same conditions, the wealth of
ancillary data, and the rich spectroscopic follow-up comprising
multiple CO transitions to evaluate the molecular gas conditions via
line luminosity ratios and average full SLEDs.  
In addition, we have access to
well-sampled far-IR dust spectral energy distributions (SEDs) and
neutral atomic carbon lines (\ci), two alternative proxies of the cold
ISM that we use to complement and cross-validate the results from
the classical CO proxy \citep{magdis_2012, scoville_2016,
  papadopoulos_2004, walter_2011}. As a matter of fact, this work
cannot directly
address the spatial scale problem mentioned above due to the limited
spatial resolution of the ALMA observations. 
However, we aim to overcome the lack of systematic
  observations of direct molecular gas tracers in average
star-forming galaxies close to the peak of the cosmic SFR history
that are normally not
present in the literature, rather than specifically targeting bright
QSOs that have been well covered across redshifts in the past. The number statistics of our sample are also
the base to determine average trends and, critically, their
dispersion. 

We briefly recall the specifications of our ALMA survey in the coming
sections, but we refer the reader interested in the details of the
observations to V20a for a complete description. Here we focus on
basic correlations among observables or directly derived quantities
and comment on their physical meaning in the various sections and
in the final discussion. The master catalog containing all the
measurements necessary to reproduce this analysis is publicly
available (V20a).
For the sake of reproducibility of the results in this work, we report
all the data used here, including updates, in a dedicated table
available in electronic
format\footnote{\label{footnote}\url{https://cdsarc.unistra.fr/viz-bin/cat/J/A+A/XXX/XXX}}. The 
data products are described in Table \ref{tab:datatable} in Appendix \ref{app:table}.
Unless stated otherwise, we assume a Lambda cold dark matter cosmology with
$\Omega_{\rm m} =0.3$, $\Omega_{\rm \Lambda} =0.7$, and $H_0 = 70$ km
s$^{-1}$ Mpc$^{-1}$ and a Chabrier initial mass function
\citep[IMF;][]{chabrier_2003}. All the
literature data have been homogenized with our conventions when necessary. 
\begin{figure*}
  \centering
  \includegraphics[width=\textwidth]{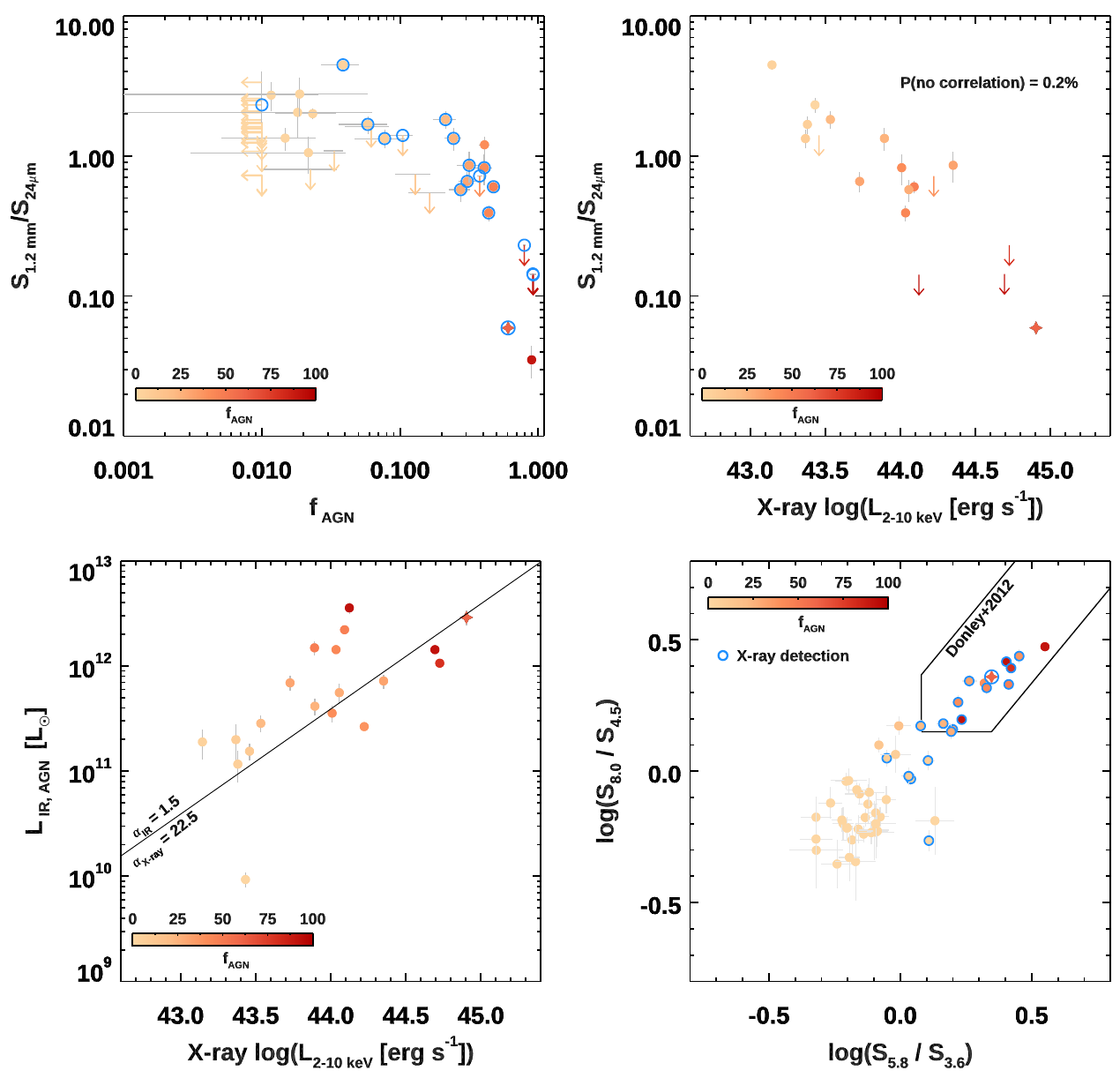}
  \caption{{How observed colors, the full far-IR SED
      modeling, and X-ray detections compare as tracers of the AGN activity.}
    \textit{Top:} Observed $S_{\rm 1.2\,mm}/S_{\rm 24\,\mu m}$ color
    (Rayleigh-Jeans/mid-IR rest-frame color as a function of \fagn\ and the X-ray
    luminosity $L_{\rm 2-10\, keV}$. \textit{Bottom left:} AGN IR luminosity from
    SED modeling as a function of $L_{\rm 2-10\, keV}$. The black line indicates
    the locus of constant bolometric luminosities derived from X-ray
    and IR luminosities adopting the corrections as labeled. \textit{Bottom
      right:} \textit{Spitzer}/IRAC [8.0]-[4.5], [5.8]-[3.6] $\mu$m color
    selection as in \cite{donley_2012}. The symbols indicate our sample
    of $z\sim1.2$ galaxies color coded by \fagn \ as in previous figures. Open blue circles
    mark hard X-ray emitters with $L_{2-10\,\mathrm{keV}}>10^{43}$
    \es. In the top panels, arrows indicate $3\sigma$ upper limits on the ALMA continuum
    emission at 1.2 mm.}
  \label{fig:fig2}
\end{figure*}

\section{Data}
\label{sec:data}

\subsection{ALMA observations and far-IR modeling}
 We extensively presented our sample and ALMA survey\footnote{Programs
   \#2015.1.00260.S, \#2016.1.00171.S (PI: Daddi), \#2018.1.00635.S,
   \#2016.1.01040.S, and \#2019.1.01702.S. (PI: Valentino).} in V20a and
we refer the reader to that work for further details. Briefly, we
selected IR-detected galaxies in the COSMOS field
\citep{scoville_2007} on and above the main sequence for a series of
subsequent follow-up observations of several CO ($J=2,4,5,7$) and \ci\ lines
(\cione, \citwo) with ALMA. We imposed a detection in
\textit{Herschel}/PACS 100 or 160 $\mu$m from the PACS Evolutionary
Probe survey
\citep[PEP;][]{lutz_2011}. This requirement privileged the selection of
massive (median $M_{\star}=10^{10.7}\,M_\odot$) upper
main-sequence sources ($L_{\rm IR}>5\times10^{11}$ \lsun) and warm
dust temperatures at a median redshift of $z\sim1.2$.
Here we focus on the sample of 55 galaxies with a robust
spectroscopic redshift from submillimeter lines (at least one line with
$Flag=1$ in V20a) and an estimate of the IR luminosity from the
modeling of the SED. We note here that the ALMA observations were
carried out at a spatial resolution of $0.8-1.4\arcsec$ (Table 1 in V20a), which was
sufficient to estimate global galaxy properties and derive reliable FIR sizes at high S/N ratios
\citep{puglisi_2021}, but not to highly resolve the central regions
of the galaxies.

The modeling of the SED was
performed on the de-blended
far-IR catalog by \citet{jin_2018} and our ALMA continuum
emission estimates at rest-frame $\sim370,\, 520,\, 610$ $\mu$m, and
$1.3$ mm when available (Fig. \ref{fig:fig1}). We excluded sources without an entry in
  the far-IR catalog (see V20a).
We used the customized $\chi^2$ minimization tool \textsc{Stardust}
\citep{kokorev_2021}\footnote{\url{https://github.com/VasilyKokorev/stardust}}
that includes representative templates spanning the expanded library of
\citet{draine_2007} and AGN models from
\citet{mullaney_2011}. The algorithm returns the total IR luminosity integrated
between 8 and 1000 $\mu$m (\lir) split into a star-forming (\lirsfr,
associated with the \citealt{draine_2007} templates) and
an AGN (\liragn) components ($L_{\rm IR} = L_{\rm IR,\,SFR}+L_{\rm
  IR,\,AGN}$). We define the AGN contribution to the
total IR budget as the fraction $f_{\rm AGN}=L_{\rm
  IR,\,AGN}/L_{\rm IR}$.

We caution the reader against low \fagn\ at face value. Fractions
$<1$\% simply indicate the absence of
a significant AGN contribution at long wavelengths, given the available
data, and we adopted this threshold as a strict floor for our
analysis. Values on the order of $\sim 5-10$\% partially overlap with hard
X-ray detections and they are, thus,
more reliable for an AGN classification in the relative scale set
  by our modeling (Figs.
\ref{fig:fig2} and \ref{fig:app:fagnxray}, 
Sect.
\ref{sec:effectsed}). We note that setting a more conservative minimum threshold of
$f_{\rm AGN}>10$\% does not affect the main conclusions of this
work. We retained the best-fit values for \fagn\ and their uncertainties
in order to allow for a continuous variation in the AGN contribution
to the IR energy output of their host, but we stress that the
exact values -- especially at low \fagn\ -- sensitively depends on the
choice of the AGN templates, assumptions in the fitting procedures,
and photometric quality and coverage, and should be thus taken with a
grain of salt (e.g., \citealt{ciesla_2015}). For reference, the \fagn\
from our decomposition are consistent with the reanalysis of the full
UV-to-radio SED modeling of part of our sample with \textsc{Michi2} (see \citealt{liu_2021} and
\citealt{kokorev_2021} for the performances of the two codes and a
broader comparison of the derived properties with alternative modeling
schemes, e.g., \textsc{Cigale}, \citealt{noll_2009}).

Conversely, large \fagn\ values of $\gtrsim80$\% mostly occur
for galaxies with upper limits at observed wavelengths longer
than $24$ or $100$ $\mu$m. In these cases, the standard modeling
  with all parameters left free to vary initially hit the boundary of the coldest
  template allowed. Therefore, in order to put physically
meaningful limits on \lirsfr\ and derive the final \fagn\ shown in
  Fig. \ref{fig:fig1}, we refit the SEDs with
\cite{draine_2007} templates with a fixed intensity of the radiation field heating the
dust (\umean). This is a unit-less
quantity that can be expressed as $\langle U \rangle \approx 1/125\times
L_{\rm IR,\,SFR}/M_{\rm dust}$, where the constant indicates the power absorbed per unit
dust mass \mdust\ in a radiation field with $\langle U \rangle = 1$
\citep{draine_2007,magdis_2012}. In the 4/55 cases with extreme \fagn\ (Fig.
  \ref{fig:fig1}), we thus fixed \umean\ to the average value for galaxies at the same
redshift and we allowed 
for a 30\% variation, comparable with the observed scatter of the
$z$-\umean\ relation
($\langle U \rangle = 15\pm 5$, \citealt{magdis_2017}). 
We note that the starburst-like
\umean\ values allowed by the current constraints would decrease
\lirsfr. This procedure mostly returns \lirsfr\ consistent with SFR of
main-sequence galaxies.

The code also outputs the dust mass \mdust, directly proportional to
the luminosity in the Rayleigh-Jeans
tail of the SED, and the aforementioned
\umean, which can be regarded as a proxy for the (mass-weighted) dust
temperature. We note that \umean\ is associated with
the \cite{draine_2007} templates and it accounts for the heating
from the star-forming component only. When necessary, we recomputed a
posteriori the total \umean\ including the AGN component by
substituting \lirsfr\ with \lir\ in the equation above. The
uncertainties on the best-fit parameters are computed as the 16th-84th
percentile confidence intervals over 10,000 randomized realizations of the
fitting process. We show the compilation of best-fit models
and the photometry for the 55 galaxies considered here in Fig.
\ref{fig:fig1}, color coded according to \fagn\ and the X-ray detection.
All the best-fit
parameters (footnote \ref{footnote}) and plots of individual
SEDs\footnote{\url{https://doi.org/10.5281/zenodo.3967380}} are available online. 
Finally, we note that one object in our survey parent sample (\#51280)
has been analyzed in detail by \cite{brusa_2018}, who collected deeper
observations of the \cofive\ emission and its underlying continuum. We
included their ALMA measurements and stellar mass estimate, but
retained our far-IR SED modeling for consistency with the rest
of the sample.

\subsection{Ancillary data}
The X-ray to radio photometry in the COSMOS field
complements the submillimeter spectroscopy with ALMA and allows for
the derivation of reliable stellar masses \citep{laigle_2016,
  muzzin_2013}. We checked for possible X-ray detections in the 2-10 keV band
with \textit{Chandra} \citep{marchesi_2016, civano_2016} and fixed a
minimum threshold of $L_{\rm 2-10\, keV}=10^{43}$ \es\ for a
significant AGN detection, which we found in 18/55 sources. We do not
retrieve any detections below this limit in our sample. Such a threshold
corresponds to SFRs $0.5-3.6$ and $2.3-5.9$ dex larger
than those we estimate from \lirsfr\ following
\cite{ranalli_2003} and \cite{lehmer_2010}, respectively, supporting
nuclear activity as the powering source of the X-ray emission.
We also note
that the distribution of observed \lx\ in COSMOS at these redshifts
peaks above $10^{44}$ \es, below which the completeness drastically
decreases while the X-ray luminosity function of AGN keeps rising
\citep{aird_2015}. 
For objects with AGN signatures in the X-ray or mid-IR regimes,
we remodeled the UV to near-IR photometry from
\citet{laigle_2016} as in \citet{circosta_2018}, providing for a
self-consistent treatment of the AGN emission across
the short wavelength spectrum. This
improved the stellar mass estimates of a handful of bright AGN hosts, while not
introducing any offsets in \mstar\ with respect to the standard COSMOS
catalogs for the rest of the sample (Fig. \ref{fig:app:mass} in Appendix).

\section{Analysis and results}
In this section we present the main findings of 
  this work in terms of observables or directly derived
  quantities, such as dust emission, gas line luminosities, stellar
  masses, and SFRs. Their physical interpretation and insertion in the
  current research landscape will be further discussed in Sect.\ref{sec:discussion}.
\begin{figure*}
  \centering
  \includegraphics[width=0.9\textwidth]{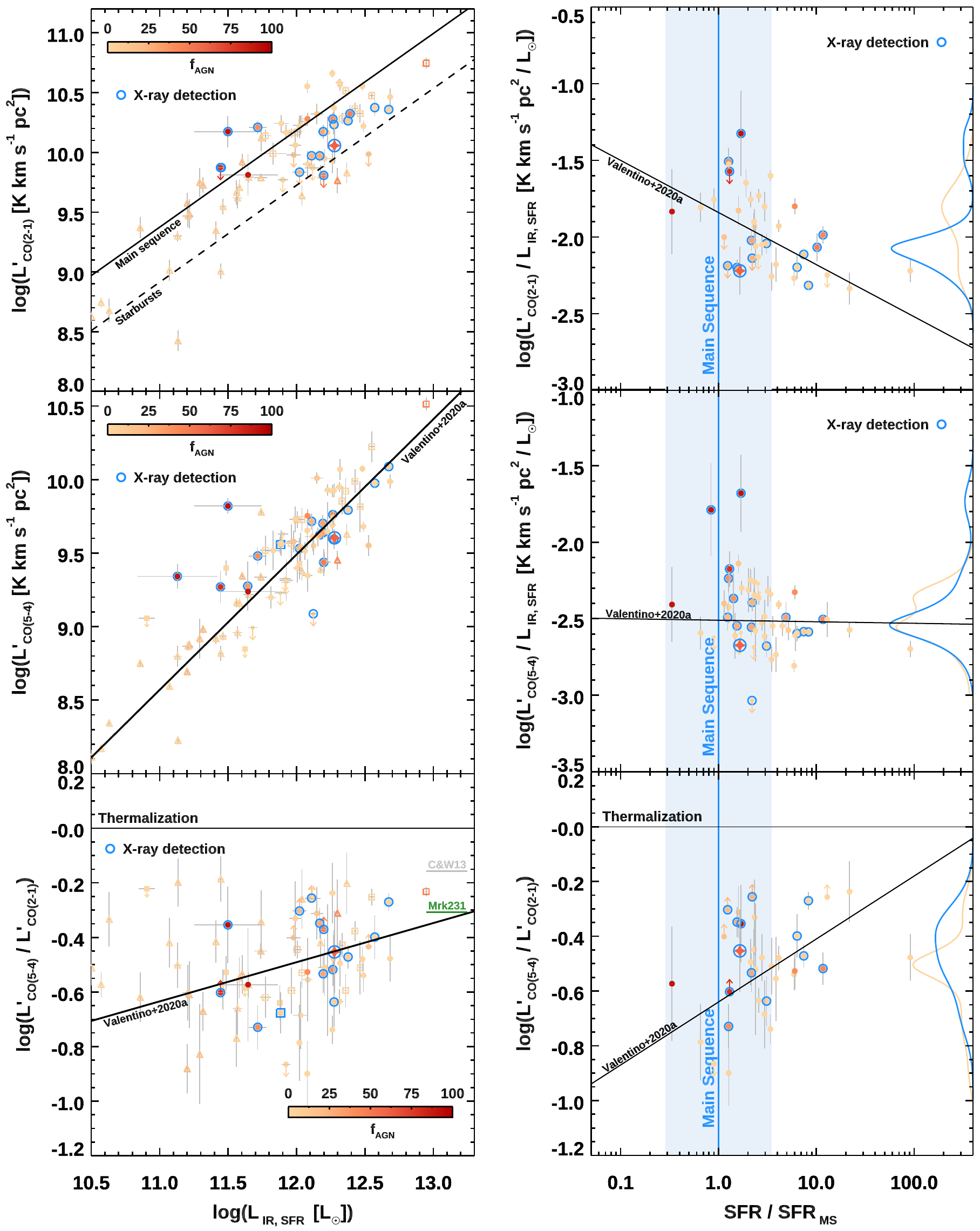}
  \caption{{AGN effect on CO line emission.} \textit{Left column,
      from top to bottom:} \cotwo, \cofive\ luminosities and their ratio as a
    function of the IR luminosities from the star-forming component
    \lirsfr. \textit{Right column:} \cotwo/\lirsfr, \cofive/\lirsfr, and
    \cotwo/\cofive\ luminosity ratios as a function of the distance from the main sequence as
    parameterized in \cite{sargent_2014} (blue area and vertical solid line).
    Filled circles mark our sample at $z\sim1.2$ color coded
    according to \fagn. Empty blue circles show hard X-ray
    emitters with $L_{2-10\,\mathrm{keV}}>10^{43}$ \es\ detected by
    \textit{Chandra}. The orange filled star marks the object from
    \cite{brusa_2018}, which is part of our parent sample. Open triangles and squares
    represent local objects and distant SMGs from \cite{liu_2021}
    color coded according to \fagn. Filled squares indicate the ASPECS
    sample, with X-ray AGN marked with blue empty squares \citep{boogaard_2020}.
    The gray and green dashes mark the
    \ratio\ for the average SLED of distant QSOs in \cite{carilli_2013}
    and Mrk 231 \citep{vanderwerf_2010}, respectively. Solid
black    lines show the loci of main-sequence
    and starburst galaxies in \cite{sargent_2014} (top left panel) or
    the best-fit relations in V20a, as labeled. The solid blue and yellow
lines in the right column show the kernel density estimation
    (KDE) of the properties on
    the Y axis for \textit{Chandra}-detected AGN and star-forming
    galaxies normalized to their relative frequency and rescaled for
    visibility. Only detections
    are included in the KDE.}
  \label{fig:fig3}
\end{figure*}
\subsection{Widespread AGN activity affects the dust far-IR emission of distant galaxies}
\label{sec:effectsed}
The far-IR SED modeling reveals the presence of common AGN
activity in our sample of $z\sim1.2$ galaxies, as anticipated in
V20a. Thirty percent of the parent surveyed
sample has $f_{\rm AGN}+1\sigma_{f_{\rm AGN}}\geq20$\%, thus
indicating a relevant contribution of the AGN template in the SED
modeling, in agreement with previous reports
\citep{mullaney_2012}. We find the same proportion
also for the subsample considered
here, thus arguing against any trivial selection bias in the line
detectability. We also retrieve 6/55 objects with $f_{\rm
  AGN}+1\sigma_{f_{\rm AGN}}>50$\%. For reference, we use
  the 20\% threshold to broadly classify galaxies as AGN based on
  their IR emission, but the
  following analysis mostly relies on the continuous distribution of
  \fagn\ values and its relation to the properties of the galaxy
  hosts.\\

We stress once again that the possibility to detect the mid-IR emission
from AGN depends on the coverage -- here granted by
\textit{Spitzer}/MIPS photometry -- and the intrinsic brightness of
the dusty tori and dust heated by the AGN compared with the underlying
emission powered by young stars. Therefore, given the positive correlation between stellar masses
and SFR, the mild skewness of the \mstar\ distribution of AGN hosts
toward low values is not completely surprising (Fig.
\ref{fig:app:mass}): at fixed \liragn, it is easier to detect the emission of an AGN
in lower mass galaxies where the contrast with \lirsfr\ is higher. 
In addition, the choice of the templates for the
decomposition affects \fagn\ at face value. While a direct comparison
with other works purely based on \fagn\ should be drawn with caution, we can
check our classification against alternative AGN indicators. The
information enclosed in \fagn\ is comparable with the AGN
classification based on \textit{Spitzer}/IRAC
colors (Fig. \ref{fig:fig2}). A smooth
gradient of \fagn\ enters the box hosting AGN designed by
\cite{donley_2012}, but there are X-ray-detected sources outside it.
Above $f_{\rm AGN}+1\sigma_{f_{\rm AGN}}>20$\%, a criterion based on \fagn\ largely overlaps with the
detection of hard X-ray photons in \textit{Chandra}'s 2-10 keV band
(Fig. 4 in V20a; see also \citealt{brown_2019}). Here $86$\% percent
of the sample with $f_{\rm AGN}+1\sigma_{f_{\rm AGN}}\geq20$\% has
$L_{2-10\,\mathrm{keV}}>10^{43}$ \es (not applying the modest
  $\lesssim5-15$\% absorption corrections, \citealt{marchesi_2016}). Conversely, $\sim13$\% of the 
population below this \fagn\ threshold is detected
by \textit{Chandra}, gauging the degree of overlap between these two
AGN selection criteria in a regime where star-formation dust heating
dominates the IR budget. Similar values are found for the whole survey
parent sample. We thus consider the X-ray-detected sources
  as AGN hereafter.\\

The overlap between the mid-IR and X-ray definition is reflected
on the shapes of the overall SEDs (Fig. \ref{fig:fig1}). The
presence of an AGN boosts the emission in the mid-IR and only
apparently suppresses the
luminosity in the Rayleigh-Jeans tail of the SED at fixed \lir\
due to the normalization. This
is shown in Fig. \ref{fig:fig2} in the
form of the observed $S_{\rm
  1.2\,mm}/S_{\rm 24\,\mu m}$ color ($\approx S_{\rm 500\,\mu
  m}/S_{\rm 11\,\mu m}$ rest-frame) probing two critical regimes
proportional to the
total dust mass content and the AGN energy output,
respectively. Given the small redshift range probed
  by our sample, we used the observed photometry to
  compute this color, thus being model independent.
The observed $S_{\rm
  1.2\,mm}/S_{\rm 24\,\mu m}$ color tightly anticorrelates with
\fagn\ and the $L_{2-10\,\mathrm{keV}}$ luminosity (Fig.
\ref{fig:fig2}). We note that the total luminosity density at
rest-frame 11 $\mu$m
independently correlates with \lx\ while it does not at 500 $\mu$m,
once the effect of the stellar mass is folded in the calculation. This suggests
that most of the AGN effect is exerted in the mid-IR as a result of
increased dust heating, but without any
apparent impact on the Rayleigh-Jeans tail and, thus, on the total dust
mass content, as it could be expected in a scenario with an
  efficient negative feedback onto the host (Sect.
  \ref{sec:discussion}). 

We also note that
\liragn\ correlates well with $L_{2-10\,\mathrm{keV}}$ in our sample (Fig.
\ref{fig:fig2}; see \citealt{gandhi_2009}) and supports the interchangeable usage of \fagn,
$L_{2-10\,\mathrm{keV}}$, and \liragn\ as tracers of AGN activity
in our sample. On the other hand, we do not find any significant
  correlation between \liragn\ or \lx\ and \lirsfr\ at fixed stellar
  and dust masses, proxies for the specific SFR (sSFR) and SFE, respectively. The AGN
  activity seems not to suppress nor enhance the formation of new
  stars or its efficiency in the galaxies in our sample. However, we
  note that \lirsfr\ and \lx\ probe star formation and
  AGN activities characterized by different timescales ($\sim100$ Myr and 
  $<10$ Myr, respectively). This is a known limitation to the comparison
  between these quantities \citep[e.g.,][]{hickox_2014,stanley_2015}. 

\subsection{The effect on CO line luminosities}
\label{sec:columinosities}
Once we factor out the contribution of AGN-heated dust to the total IR
luminosity ($L_{\rm IR,\,SFR} = L_{\rm IR}-L_{\rm IR,\,AGN}$), the central nuclei seem to marginally affect the CO emission. In
Fig. \ref{fig:fig3}, we show the \cotwo\ and \cofive\ $L'$ luminosities
as a function of \lirsfr. The line luminosities are defined as:
\begin{equation}
L'_{\rm line}\,[\mathrm{K\,km\,s^{-1}\,pc^2}] = 3.25\times10^7\,S_{\rm
  line}\,\Delta v\, \nu_{\rm obs}^{-2}\,(1+z)^{-3}D_{\rm L}^2
,\end{equation}
where $S_{\rm line}\,\Delta v$ is the velocity-integrated line flux in
\jykms, $v_{\rm obs}$ is the observed line frequency in GHz, $z$ is
the redshift, and $D_{\rm L}$ is the luminosity distance in Mpc \citep{solomon_2005}.
For reference, we also show
the sources detected by the blind ALMA Spectroscopic Survey in
  the Hubble Ultra Deep
Field (ASPECS) at $z\sim1.2$ and $\sim2.5$ \citep{boogaard_2020}, whose SEDs we remodeled using our
algorithm, notably including \textit{Herschel} photometry (T. Wang,
private communication). Moreover, we also added the literature
compilation from \cite{liu_2021} who applied an SED modeling similar to
and consistent with ours to nearby
galaxies and distant SMGs. For a meaningful comparison,
we show objects in the same range of redshift and luminosities of our
sample. Including local galaxies with lower luminosities and the handful of powerful (possibly
lensed) SMGs at $z>4$ in the compilation does not
appreciably affect the main points of this work.\\

\begin{figure}
\includegraphics[width=\columnwidth]{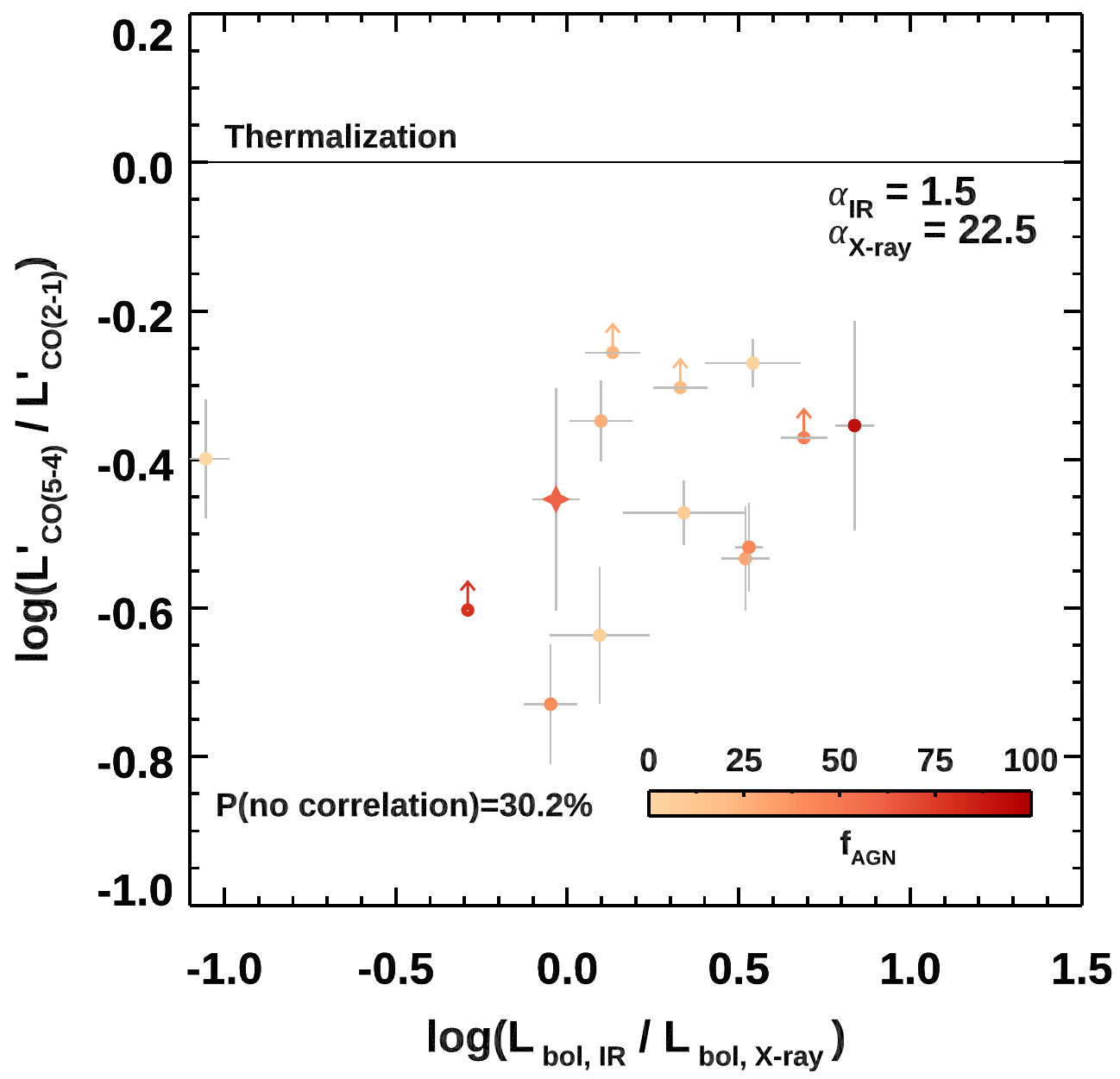}
\caption{{CO line excitation (\ratio) as a function of a proxy for heavy
    obscuration (\liragn/\lx).} \liragn\ and \lx\ have both been
corrected to the total bolometric output by factors of $1.5$ and
$22.5,$ respectively. The symbols represent our sample at
$z\sim1.2$ and are color coded as in the previous figures.}
\label{fig:agnobsc}
\end{figure}

Although with a substantial scatter, AGN hosts appear to
preferentially occupy an intermediate locus between
main-sequence and starbursting galaxies in the \lirsfr-\lprimecotwo\
plane, according to the parameterization of \cite{sargent_2014}.
When a mid-$J$ transition such as \cofive\ is
considered, galaxies hosting an AGN are fully mixed with the rest of the
star-forming population, both groups following the relation presented
in V20a. This reinforces the idea that \cofive\ can be used as a
tracer of SFR independently of the galaxy type. However, we report the
elevated \lprimecofive\ luminosity for two sources with the highest
$f_{\rm AGN}$ given their \lirsfr, despite the large uncertainties on
the latter. The $J=5/2$ luminosity
ratio $R_{52}=L'_{\rm CO(5-4)}/L'_{\rm CO(2-1)}$ naturally reflects these trends and it does not appear substantially
affected by the presence of AGN at fixed \lirsfr.
A similar conclusion is reached
comparing \ratio\ with \liragn\ or \fagn\ \citep[see also][]{liu_2021}. 
In addition, despite the low number statistics, \ratio\ does not
correlate with \liragn/\lx\ (Fig.
\ref{fig:agnobsc}). This ratio is a proxy for
heavy obscuration in AGN approaching or in the Compton-thick regime
\citep[e.g.,][]{delmoro_2016} and it allows us to check for the
existence of a possible dependence of the CO excitation on the large
dust opacities expected in strongly starbursting AGN hosts at high
\liragn/\lx\ ratios (see also Sect. \ref{sec:dust}).
We find a probability of no correlation of 30\% from a
  generalized Kendall's tau test including the upper limits. This
    is robust against the exclusion of the lowest \liragn/\lx\ value
    ($p_\mathrm{no\,corr}=25$\%). For
simplicity, in Fig. \ref{fig:agnobsc} we show the observed
luminosities corrected to the
bolometric output by adopting constant $1.5\times$ \citep{elvis_1994} and
$22.5\times$ \citep{vasudevan_2007} factors for \liragn\ and \lx,
respectively. Using different bolometric corrections from the
literature \citep[e.g.,][]{lusso_2011} does not alter
this conclusion (see \citealt{lusso_2012,duras_2020} and references
therein for more refined luminosity-dependent corrections and their
substantial uncertainties, which are
beyond the scope of this paper).\\ 

Finally, we find that the
\coseven\ line
provides a consistent picture to that of \cofive,
despite being more uncertain due to low
number statistics as 20\% of our sample has the right frequency
coverage (see \citealt{lu_2015}, \citealt{liu_2015}, and V20a for a comparison
  between \cofive\ and \coseven\ as SFR tracers).

\subsection{A standard metric: The distance from the main sequence}
\label{sec:distancems}

Marginally lower \lprimecotwo\ luminosities at fixed \lirsfr\ could potentially
suggest a prevalence of starburst-like high SFEs
($\mathrm{SFE}=\mathrm{SFR}/M_{\rm gas}$) or equivalently shorter depletion
timescales ($\tau_{\rm depl}=\mathrm{SFE}^{-1}$) in AGN
hosts. To test for this, we plot the \lprimecotwo/\lirsfr\ ratio
  of our sample as a function of the
  distance from the main sequence
  ($\Delta\mathrm{MS}=\mathrm{SFR/SFR_{MS}}$, Fig.
\ref{fig:fig3}). AGN hosts occupy the lower
end of the distribution of the ratios over the same
interval of $\Delta$MS spanned by the star-forming population (Fig.
\ref{fig:fig3}). This is also the locus occupied by the
compact objects with size measurements from ALMA \citep{puglisi_2021}. A logrank test
shows indeed a mild tendency of AGN to reside in compact galaxies,
defined based on their distance from the mass-size relation
(\citealt{puglisi_2021}; see also Sects. \ref{sec:dust} and \ref{sec:discussion}). Their reduced \cotwo\
luminosities might thus be plausibly driven by the size of the host and the
ensuing larger SFR densities, rather than the presence of the AGN. In
addition, a non-parametric 
Kolmogorov-Smirnov test returns a probability of 9.4\%
that the
\lprimecotwo/\lirsfr\ distributions for \textit{Chandra}-detected AGN
and star-forming galaxies are drawn from the same parent
population, confirming that the effect is not statistically
significant at this stage.
We note that we considered only \cotwo\ detected objects for the
calculation. 
Moreover, with the notable exception of the two objects with elevated
\lprimecofive\ luminosity mentioned in the previous section, we do not find any significant difference in
the distribution of the \cofive/\lirsfr\ or \cofive/\cotwo\ ratios as a
function of $\Delta$MS. 

\subsection{Average CO spectral line energy distributions}
\begin{figure}
\includegraphics[width=\columnwidth]{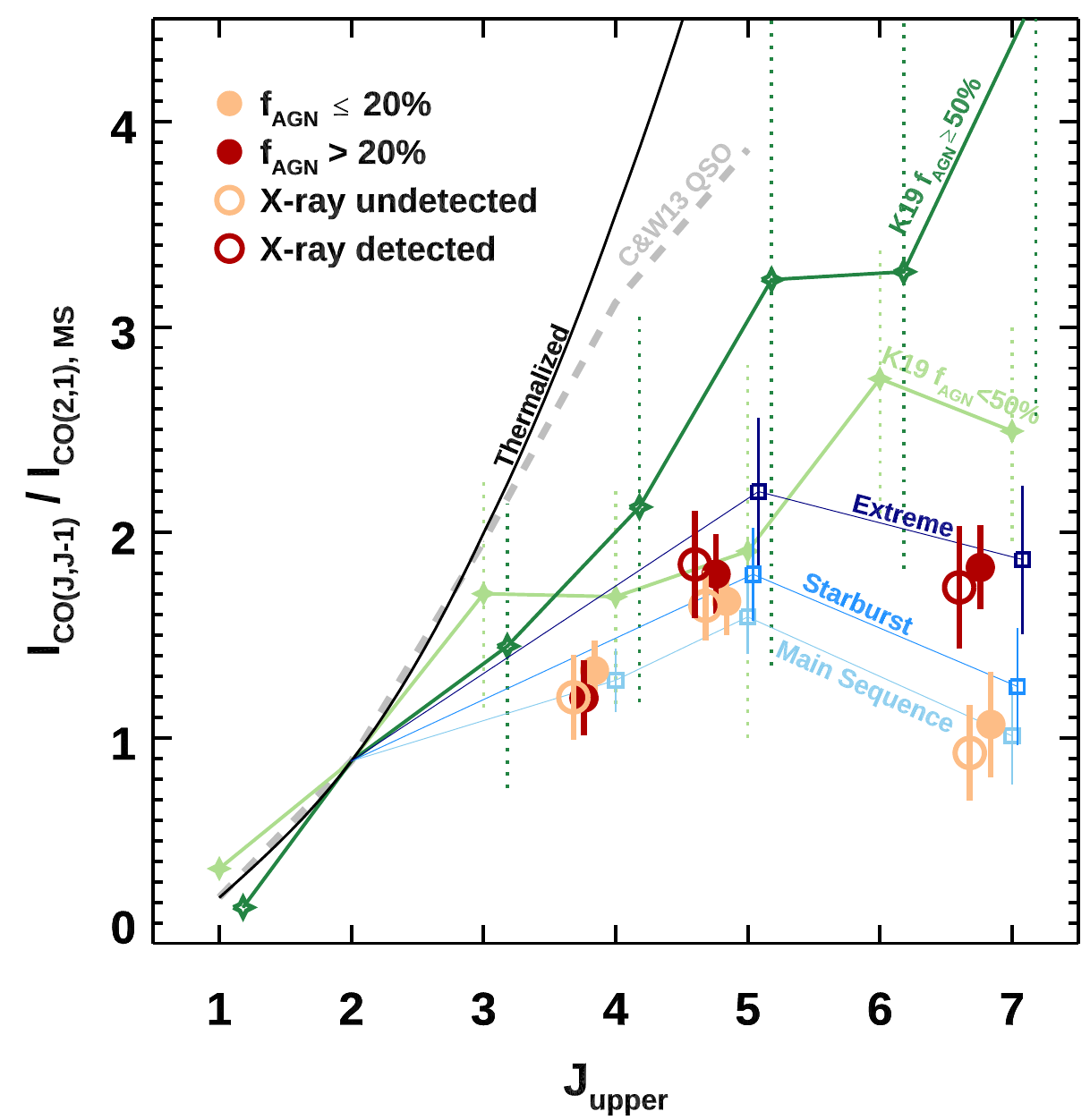}
\caption{Average CO spectral line distributions of AGN hosts and
    star-forming galaxies as well as average SLEDs for AGN hosts and star-forming
  galaxies normalized to the average \cotwo\ flux of main-sequence
  galaxies at 
  $z\sim1.2$ from V20a. Filled yellow and red
  circles indicate galaxies in our sample with $f_{\rm
      AGN}\leq20$\% and $>20$\%, respectively. Open yellow and red
  circles show objects split 
  according to their \textit{Chandra} detection at 2-10 keV. Blue open
  squares represent the SLEDs of the main sequence and (extreme) starbursts
  ($>7\times$) $>3.5\times$ above the main sequence from V20a. Filled
  light and empty dark green stars show the average SLEDs from the
  stacking in \cite{kirkpatrick_2019}, splitting AGN hosts and
  star-forming at $f_{\rm AGN,\,mid-IR}=50$\% (comparable with
    our $f_{\rm AGN}=20$\% threshold; see their Eq. 1).
  The dashed gray line marks the
  QSO SLED in \cite{carilli_2013}. The $L'$ luminosity values for our
  samples are reported in Table \ref{tab:sled}. We note that
    the statistics of X-ray detections of AGN with measurements of \cofour\ are not sufficient to derive an average estimate.}
\label{fig:fig4}
\end{figure}
 We further attempted to assess the average SLEDs of AGN hosts and star-forming galaxies (Fig.
\ref{fig:fig4}). Here we split the sample both at $f_{\rm AGN}=20$\%
(including the upper limits) and according to the
hard X-ray detection as detailed in
Sect. \ref{sec:data}. As seen above, these alternative AGN
classifications generally overlap and provide consistent
results so far. However, they do so by cutting the sample in slightly
different ways, capturing different intermediate sources,
and complementing each other when the number statistics
according to one criterion are not sufficient to compute meaningful
ratios. 
We computed the mean SLEDs using the Kaplan-Meier estimator
\citep{kaplan-meier_1958}
and we included robust upper limits ($Flag=0.5$ in V20a). The mean and
median $L'$ luminosities are reported in Table \ref{tab:sled}. The line
 ratios in Fig. \ref{fig:fig4} are expressed as fluxes at the median
 redshift of the surveyed
 sample ($z=1.25$) after averaging the luminosities to take out the
 distance effect. The average SLEDs are computed over the sources
   meeting the requirements for the selection for the \cotwo\
   follow-up in our survey, not to bias the line ratios (V20). The
 SLEDs are normalized by the value for the
 average main-sequence galaxies to ease the comparison with our
 previous work also reported in Fig. \ref{fig:fig4}. As in
V20a, we accounted for the possible bias introduced by the spectral
sampling of the \cotwo\ ALMA follow-up. This results in similar total \lir\
distributions for the objects entering the average of each line, thus
not requiring any further normalization that could bias the average
values \citep{kirkpatrick_2019}.\\

At face value, we find only a higher \coseven\ emission for 
AGN hosts than in the star-forming population. Nevertheless, the
\textit{Chandra} detected and the $f_{\rm AGN}>20$\% samples include
100\% and 25\% of the extreme starbursts
($\Delta\mathrm{MS}>7$) identified in V20a. Therefore, it does not entirely
come as a
surprise that the \coseven\ emission resembles the one of strongly
starbursting galaxies previously identified. Further splitting
  our sample according to the distance
  from the main sequence and the AGN contribution could help
  to understand the dominant mechanism setting the \coseven\
  emission, but the current number statistics
  prevent us from doing so (Table \ref{tab:sled}). We do not find any other
significant divergence in the average CO SLED between AGN hosts and
the star-forming population. This holds also when raising the $f_{\rm AGN}$
threshold to identify AGN to a more extreme 50\%. On the other hand, we do retrieve a
substantial diversity
within both subsamples (Fig. \ref{fig:individualsleds} in
  Appendix) and the large scatter likely plays a major role
in comparing single objects with statistical and -- to some extent
inhomogeneous -- samples. For reference, we also show the stacking
results for the mid-IR-selected galaxies with \textit{Spitzer}/IRS spectra from
\cite{kirkpatrick_2019} and the classical QSO sample in
\cite{carilli_2013}. These SLEDs appear more excited than the ones
presented here, likely due to their coverage of more extreme objects
with intrinsically larger total \lir\ and more powerful AGN over a
wide range of redshifts (see also \citealt{weiss_2007proceedings},
\citealt{banerji_2018}, and \citealt{bischetti_2021} for other examples of
highly excited SLEDs of hyper-luminous QSOs at $z\sim2-3$).
We stress that the primary goal of the initial survey that
  provided the parent sample for this study was to target relatively
faint IR-selected galaxies on the main sequence, normally
underrepresented in previous studies. Discounted this difference in
normalization, our conclusions are similar to
\cite{kirkpatrick_2019}, who found only marginal evidence for enhanced CO
ratios in galaxies with large $f_{\rm AGN}$, but without 
statistical significance. We note that these authors classify
  their sample according to an analogous, but not identical definition of
mid-IR $f_{\rm AGN,\,mid-IR}$ ($5-15\,\mu$m)
computed from spectral fitting or color diagrams
(for reference,
$f_{\rm AGN,\,mid-IR}=50$\% corresponds to $f_{\rm
  AGN}(8-1000\,\mu\mathrm{m})=14.75$\%, Eq. 1 in
\citealt{kirkpatrick_2019}, comparable with our fiducial
  threshold). Targeted observations with
optically selected and radio-loud AGN \citep{sharon_2016} and blind CO
searches coupled with an X-ray AGN
classification \citep{boogaard_2020} also return
analogous conclusions. This pairs up with the results on \ratio\ in Sect.
\ref{sec:columinosities} and similar findings in \cite{liu_2021}. We
return to these points in Sect. \ref{sec:discussion}.
\begin{figure}
\centering
\includegraphics[width=\columnwidth]{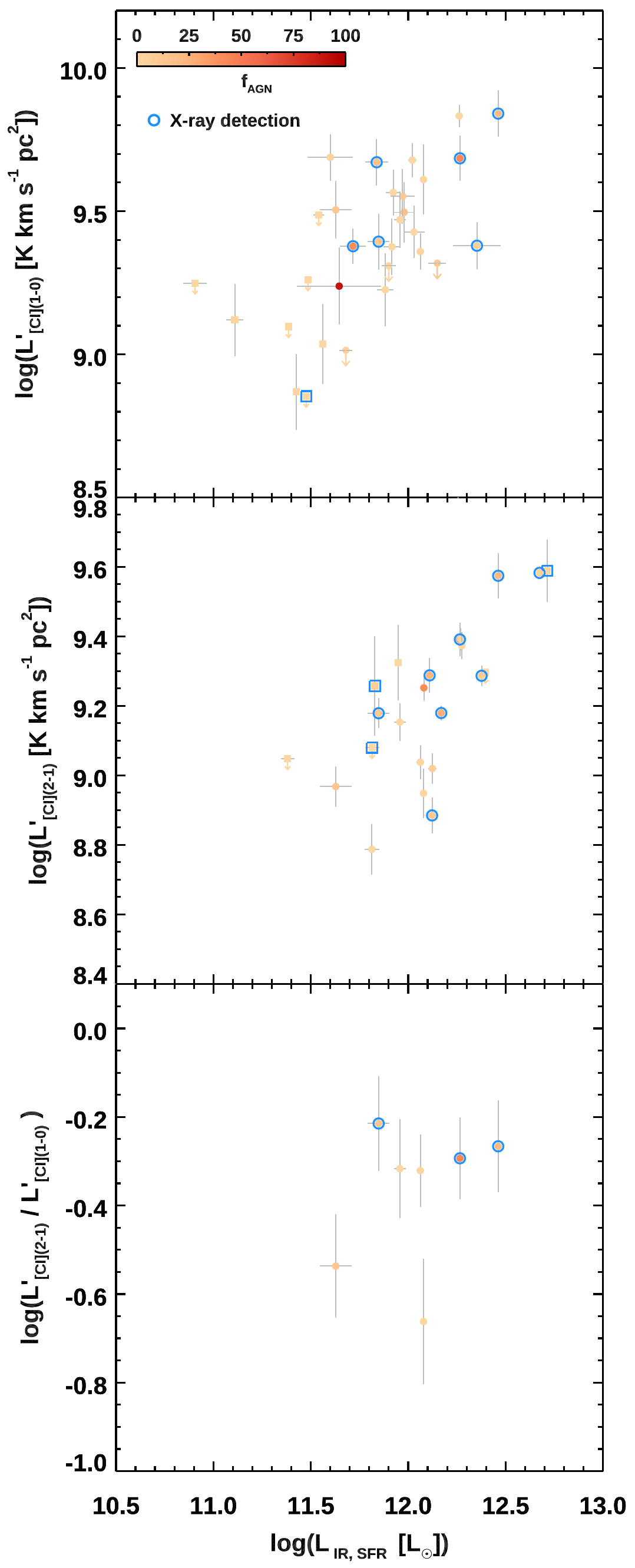}
\caption{AGN effect on \ci\ emission: \cione\ and \citwo\
  $L'$ luminosities and their ratio as a
function of the IR luminosities from the star-forming component
\lirsfr. Filled and empty blue circles mark our sample at
$z\sim1.2$ as in previous figures. Squares indicate the sample in
\cite{valentino_2018}, selected at similar redshift and luminosity.}
\label{fig:fig5}
\end{figure}

\subsection{An alternative view from neutral atomic carbon [CI] line luminosities}
Here we have the possibility to compare the results based on \cotwo\
in the previous sections with an alternative molecular gas
tracer, the neutral atomic carbon
\ci\ \citep{papadopoulos_2004, bisbas_2015, madden_2020}. Figure
\ref{fig:fig5} shows the luminosities of both \cione\ and
\citwo\ transitions and their ratio as a function of \lirsfr. We
complement our measurements with main-sequence and starburst galaxies at
$z\sim1.2$ from \cite{valentino_2018}, which were selected and modeled
similarly to the sample presented here. The statistics of objects with
large \fagn\ is scarce, thus we focus on the X-ray
classification. Based on Fig. \ref{fig:fig5}, the AGN
do not appear to have strong effects on the \ci\ emission. The
  current paucity of normal galaxies with both lines available does
  not allow us to derive a firm conclusion on the
excitation temperature as traced by the \citwo/\cione\ luminosity
ratio \citep{weiss_2005}. However, we note that the observed \ci\
ratios for both AGN hosts and star forming galaxies (SFGs) is
consistent with the average value and its large scatter for the data
compilation in \citet{valentino_2020b}. These results confirm what we anticipated in
previous works (\citealt{valentino_2018, valentino_2020b}; see also
\citealt{jiao_2017}) showing roughly constant \lprimecione/\lprimecotwo, \lprimecitwo/\lprimecione, and
\ci/dust mass ratios across not only the
IR-selected samples presented here, but also local luminous IR
galaxies, distant SMGs, and QSOs, once the effect of the AGN on the dust
continuum emission is accounted for. If any, only a mild correlation
  between the \lprimecitwo/\lprimecione ratio and the dust temperature
is present \citep{jiao_2019, valentino_2020b}. This supports the conclusions
drawn from the analysis of the low and mid-$J$ CO emission.

\subsection{A point about dust masses, opacities, and sizes}
\label{sec:dust}
\begin{figure*}
  \includegraphics[width=\textwidth]{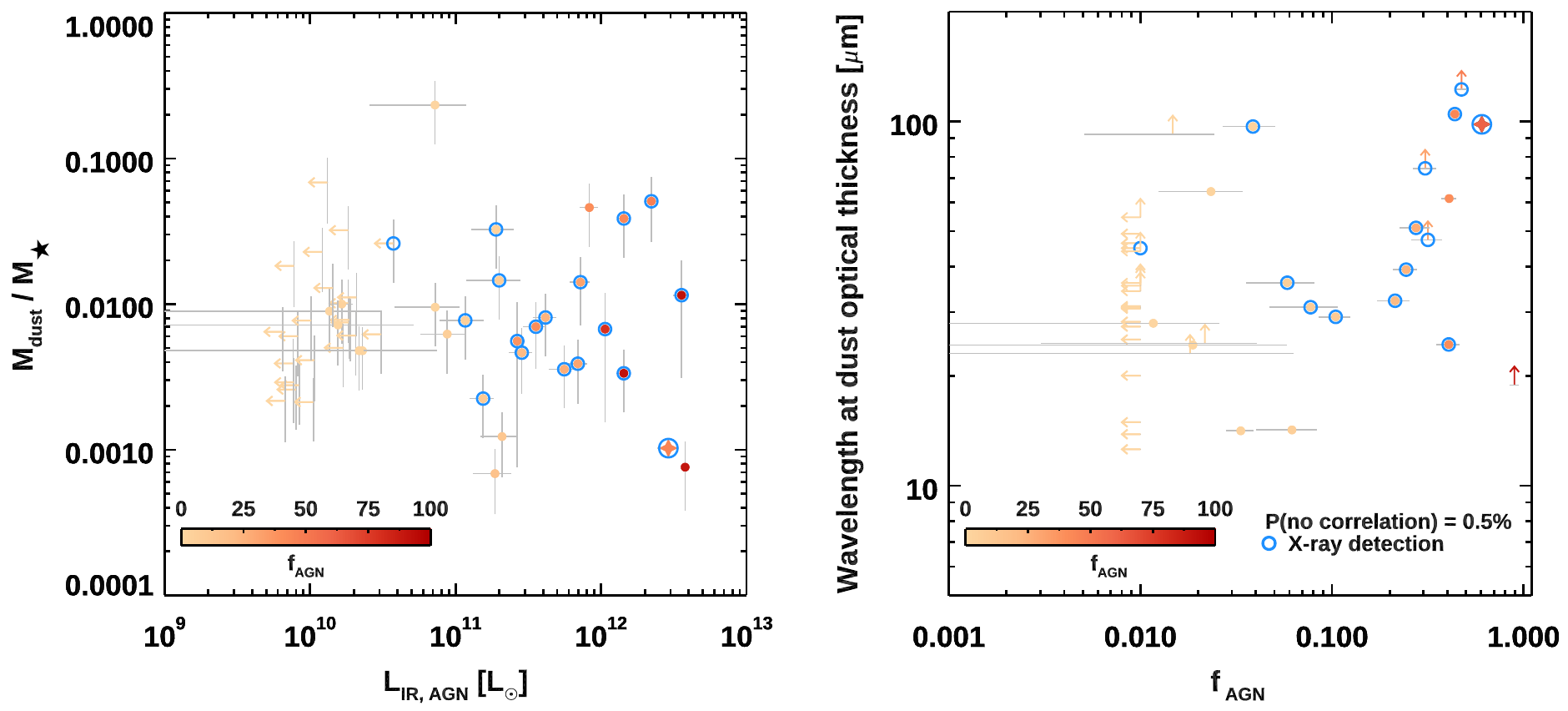}
  \caption{{Effect of AGN on the dust fraction and
      opacity}. \textit{Left:} Dust fraction \mdust/\mstar\ as a
    function of \liragn. \textit{Right:} Wavelength at which the dust
    optical depth reaches unity as a function of \fagn. Symbols and
    colors mark our sample at $z\sim 1.2$ as in previous figures.}
  \label{fig:fig6}
\end{figure*}

Continuum dust emission at long wavelengths is known to trace the
cold gas in star-forming galaxies \citep{magdis_2012,
  scoville_2016}. In this section, we use it as a third
independent method to gauge the effect of AGN on the gas reservoirs in the sample. 
The full SED modeling does not show
any significant discrepancy between the \mdust\ distributions of
star-forming galaxies and AGN hosts for similar stellar masses. The
dust mass fraction \mdust/\mstar\ does not vary as a function of \lx\
or \liragn\ (Fig. \ref{fig:fig6}): galaxies with an active nucleus
appear to have the same
amount of dust as star-forming galaxies of the same stellar mass and
location with respect to the main sequence. On the other hand, we
retrieve the increasing dust and gas fraction as a function of the
distance from the main sequence \citep[e.g.,][]{magdis_2012,
  tacconi_2018, elbaz_2018}. This mirrors the result based on the single-band
rest-frame 500 $\mu$m luminosity density (Sect. \ref{sec:effectsed}).

We note here that the \cite{draine_2007} far-IR templates used for
the SED fitting assume
that the continuum emission is optically thin at long wavelengths,
which is also the premise to use dust as a cold gas
tracer, while the AGN templates are purely empirical and could
  include self-absorption \citep{mullaney_2011}. Systematically larger dust optical thickness for AGN hosts could thus
conceal truly lower dust masses and higher temperatures. This is known
to be a factor for the most extreme dusty objects across cosmic time
\citep[][and many others]{blain_2003,  spilker_2016, scoville_2017_arp220, simpson_2017,
  jin_2019, cortzen_2020}. We tested this possibility by computing a
posteriori the wavelength at which the dust optical depth $\tau_{\rm dust}$ reaches
unity. We calculated $\tau_{\rm dust} = \kappa(\nu)\,\Sigma_{d} = \kappa_{\rm
  850\,\mu m}\,( \nu/\nu_{\rm 850\,\mu m})^{\,\beta}\,M_{\rm
  dust}/(2\pi R^2)$, where $\kappa$ is the
frequency-dependent dust opacity, and $\Sigma_{d}$ the dust mass
surface density \citep[e.g.,][]{casey_2014}. We used $\kappa_{\rm
  850\,\mu m}=0.43\,\mathrm{cm^{-2}g^{-1}}$ and $\beta=2$
\citep{li_2001}, \mdust\ from the full SED modeling (Sect.
\ref{sec:data}), and the sizes or upper limits from the ALMA
measurements when available ($R=\mathrm{FWHM}/2$ from the Gaussian
source extraction, V20a). The use of the far-IR sizes breaks the
apparent tautology when using \mdust\ from the SED modeling. The same calculation with \mdust\ derived from an ideal SED
  modeling with the opacity as a free parameter would be lower,
  further decreasing the value of $\tau_{\rm dust}$. 
In Fig. \ref{fig:fig6}, we show the relation between the wavelength
$\lambda_0$ at which $\tau_{\rm dust}=1$ and \fagn. Here we limited
the analysis to galaxies with at least one
detection of the continuum emission with ALMA to avoid severe blending
issues and poor constraints on the Rayleigh-Jeans luminosities, while lower limits on
$\lambda_0$ (due to upper limits on the size) are included. In this extreme case, we find a 
correlation between \fagn\ and $\lambda_0$ (probability of no
correlation of 0.5\% from a generalized Kendall's tau test including censored data on both axes). On average, $\langle
\lambda_0 \rangle$ is $1.8\times$ longer for X-ray detected objects than the star-forming
population ($\langle \lambda_0 \rangle_{\rm X-ray} = 51\pm 8$ $\mu$m and
$\langle \lambda_0 \rangle_{\rm SFG} = 27\pm 3$ $\mu$m, difference
significant at $2.8\sigma$, Fig. \ref{fig:fig6}). However,
we note that only in a handful of cases $\lambda_0$ reaches
values close to $100$ $\mu$m, the limit above
which we assumed optical thinness in our SED modeling. This argues
against a strong bias in the determination of \mdust.
A similar conclusion is reached when comparing the distributions of
$\langle U \rangle \approx 1/125\times
L_{\rm IR}/M_{\rm dust}$ where we accounted for \liragn.
We do retrieve higher \umean\ in X-ray detected sources than
star-forming galaxies, but not significantly ($\langle U \rangle_{\rm
  X-ray} = 33.7\pm4.4$ and $\langle U \rangle_{\rm
  SFG} = 26.1\pm3.6$, respectively). In addition, the full
distributions of \umean\ and $\lambda_0$ for AGN hosts and
star-forming galaxies are consistent with being drawn from the same
parent distribution according to a Kolmogorov-Smirnov test ($p_{\langle U \rangle}=0.12$
and $p_{\lambda_0}=0.06$). This is further reinforced when removing the constraint
on the ALMA continuum detections.\\ 

\begin{figure}
\includegraphics[width=\columnwidth]{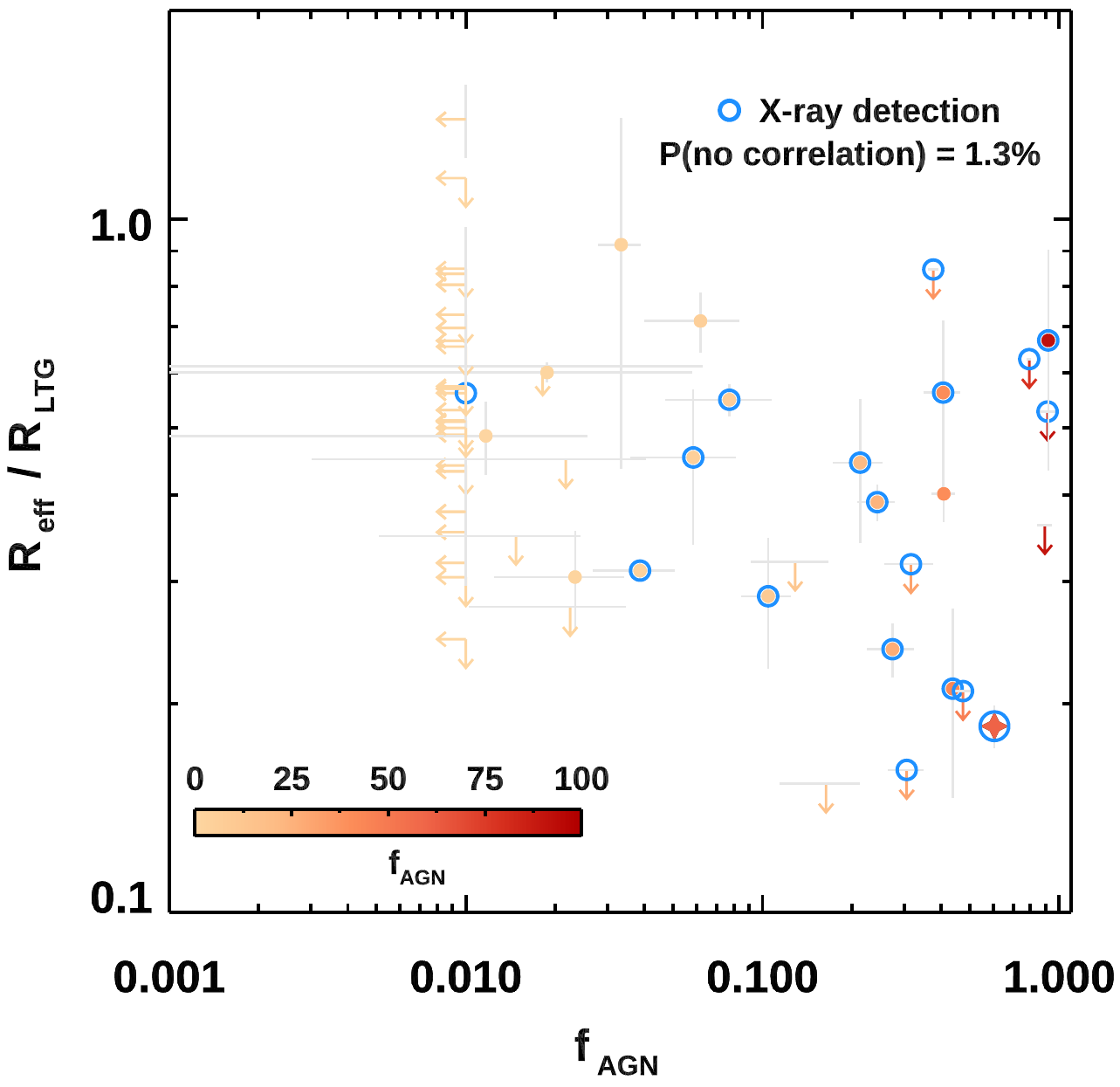}
\caption{{Compactness of AGN hosts.} 
Distance from the optical mass-size relation for late-type galaxies
expressed as $R_{\rm eff}/R_{\rm LTG}$ as parameterized in
\cite{vanderwel_2014} as a function of \fagn. The symbols represent
our sample at $z\sim1.2$ color coded according to \fagn\ as in the
previous figures.}
\label{fig:sizes}
\end{figure}

The \fagn-$\lambda_0$ relation is likely driven by a marginal
anticorrelation between \fagn\ and the host's
size at fixed \mstar\ (Fig. \ref{fig:sizes}). A Kendall's tau test
including upper limits returns a probability
of no correlation between \fagn\ and the distance from the mass-size
relation for late type galaxies at these redshifts
\citep{vanderwel_2014} of $\sim$1\%, suggesting that the compact
galaxies in our sample
tend to show enhanced nuclear activity (Sect. \ref{sec:distancems}; \citealt{puglisi_2019,
  puglisi_2021}). Nevertheless, this result is affected by the
choice of the mass-size relation necessary to anchor the calculation,
removing the effect of the stellar masses, and the
uncertainties to which it is prone. Using the compactness criterion
based on the reanalysis of far-IR and optical sizes from
\cite{puglisi_2021}, the correlation is weaker ($\sim$8\% probability
of no correlation with \fagn). The large intrinsic scatter of the
relations in place and the number statistics do not allow any stronger conclusions to be derived
at the moment.

Overall, we retrieve at best marginal differences in dust sizes, intensity of the
interstellar radiation field, and far-IR optical depth between
galaxies hosting X-ray or mid-IR detected AGN and the star-forming
population. We do not find any significant differences in dust (and
thus gas) fractions.

\section{Discussion}
\label{sec:discussion}

\subsection{Marginal effects of AGN on the average properties of
  their hosts}
The analysis of our sample of main-sequence and starburst galaxies at
$z\sim1.2$ highlights how AGN activity is distributed among star-forming
galaxies and securely detectable at IR wavelengths. A few recent literature works
push the AGN contribution to the luminosity output of the brightest QSOs
and SMGs to the submillimeter regime (\citealt{symeonidis_2016, symeonidis_2017,
  shanks_2021}). However, the debate on the effective emission at the
longest wavelengths is still ongoing (see \citealt{bernhard_2021} for a recent
assessment and detailed comparison of several AGN templates available in the literature). Overall, if not properly taken into
account, the IR emission connected with nuclei activity can cause an
overestimation of the SFRs in dusty galaxies. We reiterate that
  the initial selection criterion based on a detection at 100 or 160 $\mu$m
  biases the sample toward warmer dust temperatures and,
  thus, potentially larger fraction of AGN hosts among our
  targets. However, we do retrieve a fraction of active galaxies in
  our sample similar to previous works \citep[e.g.,][]{mullaney_2012}. Despite such
activity, for the most part AGN hosts share similar
properties with the rest of our star-forming sample. On average, we do not
retrieve any significant departure in terms of various CO and \ci\
line luminosities or their ratios, excitation conditions and SLEDs up
to $J=7$, dust and gas masses, fractions, or SFR.\\

In terms of CO line luminosities and excitation, our results are
consistent with the recent reanalysis of the literature reported in
\cite{kirkpatrick_2019}. We
  note that these authors adopted a mixed 
approach based on the spectral fitting of \textit{Spitzer}/IRS data
and a combination of IR colors to quantify the AGN emission
\citep{kirkpatrick_2017}. In broad terms, this approach 
is similar to ours (especially considering the consistent results that we obtain
when adopting \fagn\ from the full SED modeling and simpler color
cuts, Fig. \ref{fig:fig1}), but not identical. A second notable
difference is the homogeneous selection used here, rather than a
combination of literature data, and the
systematic coverage of more normal main-sequence objects typically
underrepresented in past work due to their lower
luminosities. Similar findings on the CO SLEDs are reported for
relatively large collections of galaxies at $z \sim 1-3$, either
consistently selected and observed \citep{sharon_2016, boogaard_2020}
or compiled from the literature \citep{liu_2021}. Despite suffering
from low number statistics, the
\ci\ line ratio -- a proxy for the gas temperature --,
also suggests no major impact of AGN on the cold gas, as previously
reported based on a larger literature compilation
\citep{valentino_2020b}. 

Nevertheless, highly excited SLEDs have been
frequently reported for the most luminous, frequently lensed, and
unobscured QSOs \citep[e.g.,][]{weiss_2007proceedings, carilli_2013,banerji_2018, fogasy_2020,
  bischetti_2021} and often associated with depleted gas
reservoirs in the host. The difference with our findings are the likely result of a
combination of factors. Primarily, the selection function: here we
inherited the AGN properties for a sample of IR-selected
  star-forming galaxies, rather than focusing on previously identified
  QSOs at the peak of their
power or in presence of documented massive outflows, just to mention two
popular criteria. This
resulted in lower AGN luminosities spread among the typical
population of star-forming objects ($\mathrm{log}(L_{\rm bol}/\mathrm{erg\,s^{-1}})\sim45$ against, for example, $\mathrm{log}(L_{\rm bol}/\mathrm{erg\,s^{-1}})>47.5$ in
\citealt{bischetti_2021}).
The presence of highly excited CO gas is then frequently associated
with copious X-ray photons from the AGN \citep{meijerink_2007,
  vallini_2019} or shocks induced by the
outflowing gas. However, considering that most of the information on the CO
excitation currently available in our sample comes from a mid- to low-$J$ transition
ratio such as \ratio, it is hard
to draw an unambiguous conclusion as most of the effect of these
mechanisms is expected to occur at high $J$. As a reference, a case study for
the effect of X-rays on the CO SLED as Mrk 231, also hosting abundant
multiphase outflows \citep{cicone_2012}, has an estimated 10\% contribution to the
\cofive\ luminosity
from the X-ray Dominated Regions (XDRs) component and a virtually null impact on \cotwo\
\citep{vanderwerf_2010}. The XDRs dominate over 
UV-photon-dominated regions at high $J$ ($J>9-10$), a regime still rarely
covered at high redshift (but see \citealt{gallerani_2014,
  carniani_2019, fogasy_2020}) and not accessible with our
data. In addition, X-rays, cosmic rays, turbulence, shocks, and
  mechanical heating in general can be induced by common processes
not related to AGN such as stellar winds, supernovae explosions,
mergers, or gas accretion
\citep[and many others]{meijerink_2013, lu_2014, rosenberg_2015,
  kamenetzky_2016, lu_2017, harrington_2021}. Considering the increasing SFR and merger
rates with redshift, it is not unreasonable to consider these
mechanisms as contributors to the CO excitation in distant AGN hosts. 
Adding the information on \ci\ could in principle offer another
  way to constrain the effect of X-rays from AGN, as elevated \ci/CO
  ratios due to CO dissociation have been reported at the center of nearby AGN
  hosts \citep{israel_2002, israel_2015, izumi_2020}. However, this would
  require observations able to resolve the central regions of
  galaxies, where the strongest effect is expected, and the simultaneous
  modeling of cosmic rays, also able to volumetrically break the CO
  molecules and predicted to be abundant in centrally concentrated
  starbursts \citep{papadopoulos_2004, papadopoulos_2018}.

Lower \fgas\ and shorter depletion timescales in AGN hosts than in
star-formation-dominated objects have also been recently reported for
samples of various sizes specifically selecting bright AGN
\citep{perna_2018, brusa_2018, kakkad_2017,
  bischetti_2021}, at odds with ours and similar findings in terms of
SFR and gas masses \citep[e.g.,][]{stanley_2017, schulze_2019,
  kirkpatrick_2019}. In particular, \cite{circosta_2021} reported
  a marginally significant difference between the \cothree\ emission
  of matched samples of purely star-forming galaxies and AGN at
  $z\sim2$. Interestingly, they covered a range of AGN luminosities
  partially overlapping with that of this work ($44.5<\mathrm{log}(L_{\rm
    bol}/\mathrm{erg\,s^{-1}}) <47$), rather than focusing only on
  bright QSOs, and chose to focus on observables without introducing
  any dependence on excitation corrections or gas mass conversion
  factors. Therefore, it might not come as a surprise that their
  results are somehow in between our findings and literature works.
In our case, the result on \mgas, \fgas, and SFE holds using three
different cold gas tracers, namely \cotwo,
\cione, and dust, thus excluding strong systematics linked to the
use of a particular proxy. Again,
the different selection criteria might play a significant
role in the interpretation of this outcome, as well as the ascertained presence of massive outflows
that can plausibly be responsible for substantial gas expulsion.
We do not find any strong signatures of molecular
outflows in our sample at the current sensitivity, 
excluding a couple of notable
cases (the object presented in \citealt{brusa_2018} or ID \#2299, not 
associated with the central nuclear activity, \citealt{puglisi_2021nat}). 
This might be because we are probing
quieter phases of AGN activity more spread among the bulk of the
star-forming population, rather than the pinnacle reached in bright
QSOs. However, deeper and more systematic observations will be necessary to constrain the faint
broad wings that we might expect to detect in our
sample. A mass
effect can also be in place, as most of the AGN
with high \fagn\ in our sample are hosted in lower mass galaxies
($M_{\star}\sim10^{10}$\msun, Fig. \ref{fig:app:mass}), a regime
where no major effects of the AGN presence are
generally observed \citep{circosta_2021}. \\

From a theoretical
perspective, recent high-resolution and idealized simulations support
these findings, with only the most luminous QSOs ($L_{\rm
  bol}\sim10^{47}$ \es) being able to affect the gas reservoirs of their
host. Inefficient coupling with the ISM, lower luminosities,
higher gas fractions, and outflow geometries can explain the lack of
impactful AGN feedback in distant gas rich galaxies \citep[e.g.,][]{gabor_2014, roos_2015,
  torrey_2020}. However, the range of findings from simulations is wide
and the debate is still ongoing \citep{biernacki_2018, barnes_2020,
  costa_2020}. Also, the cumulative effect of
  recurrent AGN feedback, rather than
  individual bursts on short timescales probed by the proxies that
  we adopted here, has been
  proposed as a factor affecting galaxy evolution, even if hard to constrain empirically \citep{harrison_2017}.
Using not only the existence of multiphase and
high-velocity outflows as benchmark for simulations, but also the
physical properties of the cold ISM will ease the comparison with data
in the immediate future. 
\begin{figure*}
  \includegraphics[width=\textwidth]{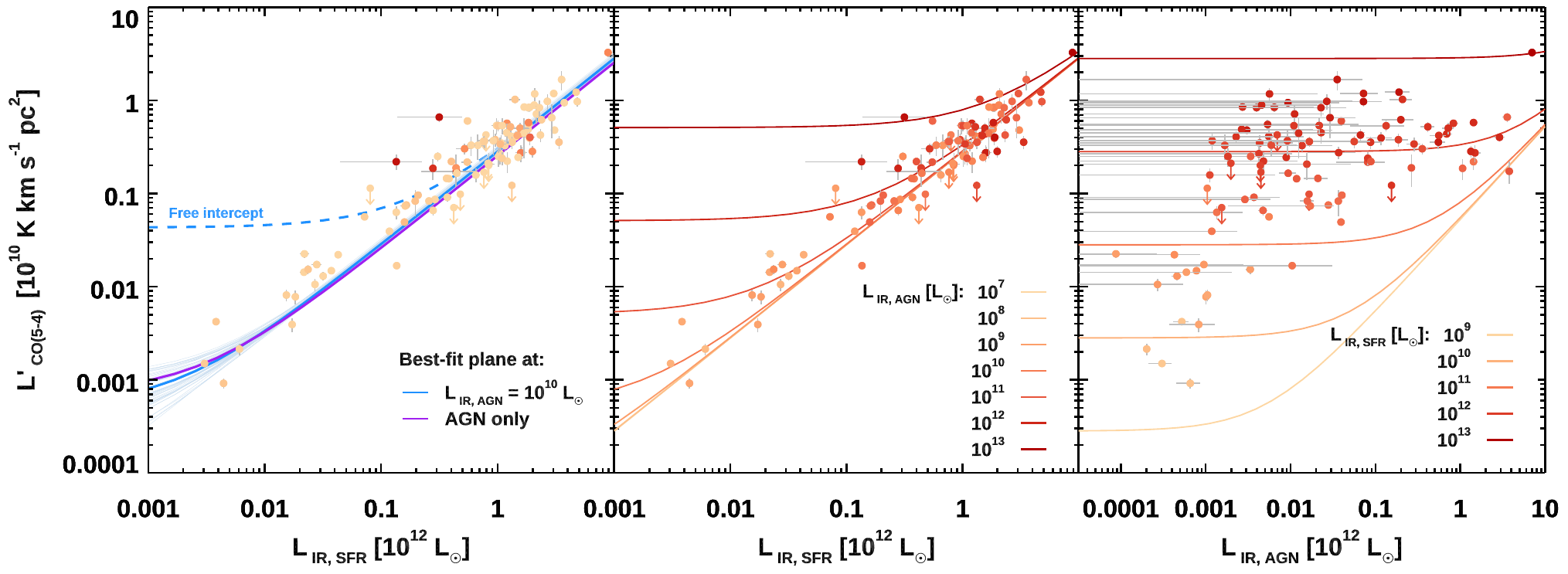}
  \caption{{\lirsfr, \liragn, and \lprimecofive\
     plane.} \textit{Left:} \lprimecofive\ as a function of
    \lirsfr. The filled circles and arrows mark the compilation of our
    measurements and $3\sigma$ upper limits at $z\sim1.2$ and from the
    literature
    \citep{boogaard_2020, liu_2021}, color coded based on \fagn. The
    thick blue lines indicate the best linear fit to
    the whole sample at the median $L_{\rm IR,\,AGN}=10^{10}$ \lsun\
    with intercept fixed to $\alpha=0$ (Eq. \ref{eq:plane}, solid line) or free to
    vary (dashed line). Thin blue lines represent random extractions
    from the posterior distribution of the best-fit model with fixed
    intercept $\alpha=0$. The purple line shows the best fit with
    fixed intercept for the sample of galaxies with $f_{\rm
      AGN}+1\sigma_{\rm f_{\rm AGN}}>20$\% or an X-ray
    detection. \textit{Center:} \lirsfr-\lprimecofive\ plane, with
    symbols marking the same sample as in the left panel but now
    color coded according to \liragn. Solid lines indicate the loci of
    the best-fit model of Eq. \ref{eq:plane} at fixed \liragn\ as
    labeled and color coded as the individual symbols. \textit{Right:}
    \liragn-\lprimecofive\ plane, with
    symbols marking the same sample as in the other panels but now
    color coded according to \lirsfr. Solid lines indicate the loci of
    the best-fit model of Eq. \ref{eq:plane} at fixed \lirsfr\ as
    labeled and color coded as the individual symbols. We report the gray error
    bars on \liragn\ to show the weighting scheme implemented for sources
    with $f_{\rm AGN}<1$\%.} 
  \label{fig:plane}
\end{figure*}

We also reported a mild tendency of AGN host to have compact
submillimeter sizes,
which could explain the larger dust optical depths compared with
the galaxies dominated by star formation. Reversing the argument, we previously reported the
presence of AGN signatures in $\sim40$\% of the compact galaxies in
our sample, classified accordingly to their distance from the
mass-size relation \citep{puglisi_2019, puglisi_2021}. It is not unusual to find
even powerful AGN and QSOs embedded in small galaxies \citep[e.g.,][]{talia_2018,
  elbaz_2018, brusa_2018,  damato_2020, bischetti_2021}. This could be
interpreted as a signature of the existence of a common mechanism triggering both the
accretion of gas onto the central SMBH, its activity, and the bursts
of star formation observed in the compact host, driving their
  coevolution. Compact hosts might also
be the preferential location where to look for the most efficient
configurations to launch massive outflows \citep[e.g.,][]{costa_2020}.
Violent gas disk instabilities, minor or major mergers, or simple gravitational
destabilization are able to funnel gas toward the central regions of
galaxies, igniting the AGN and star formation activity (V20a and
references therein). We note that the ensuing high SFR surface
densities in compact starbursting galaxies hosting an active nucleus
might be sufficient to explain the shape of the observed SLEDs at least up to
mid-$J$ transitions (V20a), as also advocated in
models and simulations \citep{narayanan_2014}, without resorting to
AGN-driven effects. This could not explain highly excited SLEDs of
QSOs found in very extended disks \citep{bischetti_2021} for which the
long reach of AGN heated photons might be a more suitable
explanation. Finally, we mention again that our observations are
  not able to resolve the central regions of the galaxies down to the
  100-500 pc scales where most of the effect of AGN is generally
  expected and observed in the local Universe
  \citep[and references therein]{vallini_2019}. This might also
  contribute to explain why more compact galaxies seem to have more
  excited gas reservoirs, as a larger fraction of molecular gas is in
  the vicinity (less than a few hundred pcs) of the active nucleus.
Future high spatial resolution measurements will be able
to disentangle the contribution of star formation and AGN on the gas
excitation.

\subsection{Nuclear activity as a factor contributing to the
  wide variety of excitation conditions in normal galaxies}
\begin{figure*}
  \includegraphics[width=\textwidth]{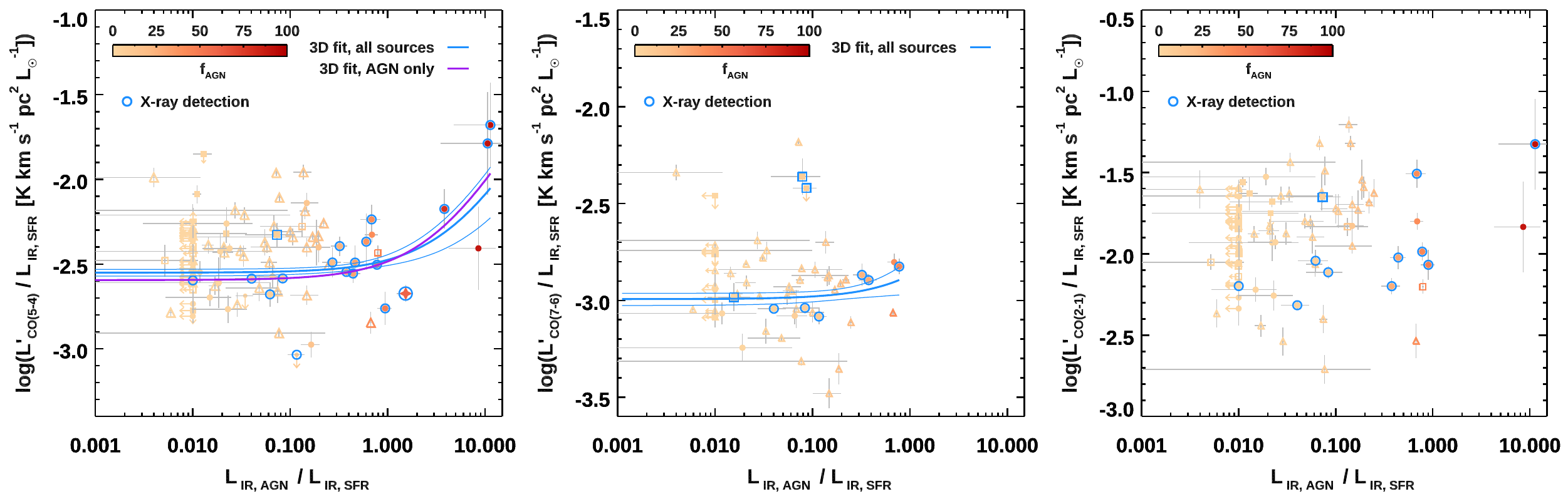}
  \caption{{CO emission as a function of the AGN-SFR contrast and}
    $L'_{\rm CO}$/\lirsfr\ ratios as a function of the contrast
    between the AGN and star formation contributions to the total IR
    budget for \cofive, \coseven, and \cotwo\ (from left to right). The
    symbols mark our sample of galaxies at $z\sim1.2$ and galaxies from the
    literature \citep{boogaard_2020, liu_2021} as in previous
    figures, color coded according to \fagn. Vertical arrows mark
    $3\sigma$ upper limits on $L'_{\rm CO}$, and horizontal arrows show
    the floor of $f_{\rm AGN}$ that we considered for the
    high-redshift galaxies. Blue open symbols indicate X-ray
    detections. The solid blue lines indicate the projection of the
    best-fit plane models with fixed null intercept in the 3D space of
    \lirsfr, \liragn, and $L'_{\rm
      CO}$ for \cofive\ (Eq. \ref{eq:plane}) and \coseven. The solid
    purple line in the left panel indicates the projection of the
    best-fit model for AGN hosts with $f_{\rm AGN} + 1\sigma_{f_{\rm
        AGN}}>20$\% or X-ray detected. The absence of points with
    $L_{\rm IR,\,AGN}/L_{\rm IR,\,SFR}>1$ in the central panel is due
    to the lack of \coseven\ coverage.} 
  \label{fig:fractions}
\end{figure*}
 If the average properties of AGN hosts and star-forming galaxies
appear consistent with each other, the observed dispersion of CO luminosities and line
ratios even for a homogeneous sample like ours still suggests an
underlying complex picture (V20a). We already mentioned that two
of the sources with the highest \fagn\ are over-luminous in \cofive\
given their \lirsfr\ (Fig. \ref{fig:fig3}) and that more powerful
QSOs are associated with highly excited CO SLEDs. This encouraged us
to investigate the possible role of AGN as a second factor regulating the
CO emission at mid to high $J$, contributing to the scatter of the
observed \lirsfr-$L'_{\rm CO}$ relations and explaining the
outliers hosting strong AGN. Given the large statistics available,
we thus attempted to model \lprimecofive\ as a linear combination of \lirsfr\
and \liragn. We included all the local and $z\sim1 - 3$ points from
our sample and the compilations in \cite{boogaard_2020} and
\cite{liu_2021}. We considered the uncertainty on every variable and
modeled the upper limits on \lprimecofive\ adopting the Bayesian
multiple linear regression code \textsc{Mlinmix\_err.pro} by
\cite{kelly_2007}, which we modified in a minor way to impose a zero constant
$\alpha$ intercept\footnote{For a comparable use of this code in 2D
  fixing a constant slope, see \cite{kilerci-eser_2015}.}. 
To account for the uncertain \liragn\ at $f_{\rm
  AGN}<1$\%, we adopted the best-fit values from the SED modeling and
boosted their uncertainties to ensure a minimal coverage of the $f_{\rm AGN}=0-1$\% range. 
The best-fit model is:
\begin{equation}
\frac{L'_{\rm CO(5-4)}}{10^{10} \mathrm{K\,km\,s^{-1}\,pc^2}} =
  (0.28\pm0.01) \times \frac{L_{\rm IR,\,SFR}}{10^{12}L_\odot} +
  (0.05\pm0.03) \times \frac{L_{\rm IR,\,AGN}}{10^{12}L_\odot} 
\label{eq:plane}
,\end{equation}
where the parameters are the median and the range enclosing the
16-84\% of the posterior probability distribution. Such a distribution
is symmetrical and adopting the mean and standard deviation
returns consistent results. We also estimate an intrinsic scatter of
  0.16 dex. We visualize this hyperplane in the 2D
projections \lirsfr-\lprimecofive\ and \liragn-\lprimecofive\ in Fig.
\ref{fig:plane}. As known, \lirsfr\ empirically correlates with
\lprimecofive\ and this is reflected also in the
\liragn-\lprimecofive\ plane where most of the gradient is due to
smoothly increasing \lirsfr\ with \liragn. This is confirmed by the
partial correlations coefficients from the fit ($p(L'_{\rm
  CO(5-4)},L_{\rm IR,\,SFR}\mid L_{\rm IR,\,AGN}) = 0.91$ and $p(L'_{\rm
  CO(5-4)},L_{\rm IR,\,AGN}\mid L_{\rm IR,\,SFR}) = 0.28$). Similar
best-fit parameters are retrieved when imposing a cut on the strength
of the AGN defined as X-ray detected or with $f_{\rm
  AGN}+1\sigma_{f_{\rm AGN}} > 20$\%. This naturally boosts the
partial correlation between \liragn\ and \lprimecofive. We note that
we adopted \lirsfr\ and \liragn\ as parameters to adhere
as close as possible to observations, but the model could be expanded
to more physical quantities as SFRs or include further corrections as
bolometric factors.\\

This simple model returns the already established
\lirsfr-\lprimecofive\ relation and its slope for $L_{\rm
  IR,\,AGN}\xrightarrow{}{} 0$, when evaluated under the
same assumptions of linearity and null intercept and over the same
sample. It is also consistent with past determinations following a
similar recipe \citep{daddi_2015}. Moreover, it decreases the
intrinsic scatter of the standard \lirsfr-\lprimecofive\ relation by 17\%, while
explaining the presence of the outliers as due to over-luminous AGN at
fixed \lirsfr. This suggests that the strength of an AGN and its contrast
with the underlying star formation activity can act as a secondary
factor contributing to the emission of a mid-$J$ CO emission as
$J=5$, for example via XDR,
and might be responsible for strong outliers with large \fagn. In Fig.
\ref{fig:fractions}, the boost of
\lprimecofive\ compared to the level expected from \lirsfr\ is shown
as a function of the AGN/SF contrast expressed in terms of
\liragn/\lirsfr, along with the corresponding projections of the plane
in Eq. \ref{eq:plane}. We note that, by modeling the data
  directly in the
  \liragn/\lirsfr\ -- \lprimecofive/\lirsfr\ space, we find a
  1.4$\times$ higher normalization for AGN hosts than when
  fitting the whole sample ($2.7\sigma$
  significance). Larger samples will allow us to test the existence,
  the robustness, and the possible physical origin
  of this tentative difference (currently not significant in the full fit in the 3D
  space, which does not suffer from the intrinsic correlation between
  \liragn/\lirsfr\ and \lprimecofive/\lirsfr). For reference, we also show the cases of \coseven\
and \cotwo. We attempted a similar parameterization for the former,
but the lack of coverage of high-contrast \liragn/\lirsfr\ sources
returns results fully consistent with the known
\lirsfr-\lprimecoseven\ relation \citep[e.g.,][V20a]{greve_2014,
  liu_2015} and no reductions in the intrinsic scatter. We do not
attempt any fit for \cotwo, given the known sublinear dependence on
\lirsfr\ and its larger scatter or possible binomial distribution \citep{sargent_2014}.\\

The intended simplicity of our empirical prescription does not exhaust
all possible explanations for the intrinsic scatter of the
\lprimecofive-\lirsfr\ relation, of course. Moreover, it does rely
on specific assumptions, mainly the linear dependence on \lirsfr\ and
\liragn\ and a fixed null intercept. Small deviations from linearity
of the \lirsfr-\lprimecofive\ relations have been indeed reported
\citep[V20a]{liu_2015}. Fixing the intercept to zero is physically
reasonable considering star formation and nuclear activity as the main
direct (e.g., supernovae explosions, cosmic rays, X-rays) and indirect
(e.g., shocks, turbulence) heating sources in distant massive
galaxies. However, it is formally justified only with the certainty that
the linear model is correct. We, thus, repeated the fitting procedure
allowing for a free intercept obtaining $\alpha=0.04\pm0.02$, which
does not significantly deviate from our physically motivated choice of
$\alpha=0$. If any, this intercept could be interpreted as a gauge of
the validity of our assumption on the linearity. Finally, we remind the reader
that the SED decomposition in presence of very luminous AGN, but scarce
photometric coverage, is uncertain. Decisive improvement will come with
mid-IR spectroscopy with \textit{James Webb James Telescope} and with
systematic observations of galaxies with well determined SEDs. Probing
the regime of high \fagn\ and spanning a large \lirsfr\ interval will
test our suggested relation and possibly extend it to higher-$J$ CO
transitions, where the effect of AGN is expected to be more significant.

\section{Conclusions}
We investigated the impact of AGN activity on the molecular gas and
dust reservoirs in a sample of IR-selected galaxies on and above
the main sequence
at $z\sim1.2$. This is part of our ALMA campaign targeting multiple
\co\ and \ci\ transitions being carried out in order to offer a complete view of the gas
masses and properties in galaxies representative of the bulk of the
star-forming population approaching the peak of the cosmic star
formation history. 

\begin{enumerate}

\item We find that $\sim30$\% of our IR-selected sample shows
  signatures of AGN activity as traced by a detailed decomposition of
  their SED at long wavelengths, an empirical color probing the mid-IR
  and the Rayleigh-Jeans regimes, and hard X-ray photon detections
  with \textit{Chandra}. 

\item The contribution of the AGN dust heating (\fagn = \liragn/\lir) represents an important
  item in the total IR budget of a galaxy, and neglecting it 
  can lead to severe overestimates of the SFR. In our sample, this is due to the excess
  power in the mid-IR portion of the SED, while we detect only very
  minor effects on its Rayleigh-Jeans tail. We do not find any
  significant suppression or enhancement of the SFR nor of the dust --
  and, thus, gas -- masses in AGN hosts at fixed stellar masses. 

\item The previous point is confirmed by the \cotwo\ and \ci\ line
  luminosities -- proxies for the molecular gas mass -- of AGN hosts, which do not
  show any statistically significant departure from the general population of
  star-forming galaxies at similar redshifts and stellar masses. We do
  not retrieve any significant trend with the distance from the main
  sequence.

\item We also reach similar conclusions for mid-$J$ \cofive\ and \coseven\ transitions
  and for the excitation conditions as probed by the \ratio\ ratio and
  the average CO SLEDs, for which we do not see any appreciable
  difference between star-formation-dominated galaxies and AGN
  hosts in general. Heavy AGN obscuration as traced by the
  \liragn/\lx\ ratio does not seem to play any role in
  determining the \ratio\ ratio either. In addition, we do not see any effect on
  the \ci\ ratio tracing the gas kinetic temperature, but with low
  number statistics. However,
  we do retrieve a wide intrinsic variety of gas conditions as probed by the
  large dispersion of the \ratio\ ratio and SLED shapes. Moreover, we
  do find a handful of AGN hosts with large \fagn\ that are potentially
  over-luminous in \cofive\ given their \lirsfr.

\item We attempted to model the \cofive\ emission as a linear
  combination of \lirsfr\ and \liragn, finding it capable of explaining
  the few outliers of the \lprimecofive-\lirsfr\ relation with large \fagn\ and reducing its intrinsic scatter. Systematic coverage of sources with
  well-sampled SEDs and a high contrast between the AGN-driven and
  star-formation-driven IR emission
  will validate or discourage the use of \liragn\ or an analogous
  metric of the nuclear activity as secondary parameters to interpret
  the mid-$J$ and higher CO transitions. 

\item We report a positive correlation between the \fagn\ and the
  wavelength at which the dust emission becomes optically thick,
  $\lambda_0(\tau=1)$ (2.8$\sigma$ significance). The dust in galaxies hosting IR
  bright AGN thus reaches the optical thickness at longer wavelengths,
  but it rarely overcomes $\lambda_0=100$ $\mu$m in our sample,
  guaranteeing robust dust mass estimates. This correlation is most likely
  driven by the mild tendency of AGN hosts to be compact, but more
  definitive results will come with larger samples.

\item We conclude that a widespread AGN activity is present in
  galaxies representative of the typical star-forming population and
  starbursts at $z\sim1.2$, but without any apparent major global effects on
  the SFR and gas or dust properties in the general population,
  with the exception of a handful of extreme cases. This might imply a common
  triggering mechanism for the star formation and AGN activity, but
  there is no obvious feedback stimulating or suppressing galaxy growth at
  this level. Our findings are partially at odds
  with previous studies that specifically focused on the brightest QSOs --
  often lensed and less obscured, and with evidence of massive
  outflows -- for which lower \fgas, higher SFEs, and more excited gas
  have been reported. However, our results are consistent with past
  works that targeted broader star-forming and SMG populations, despite the lack of
  coverage of more typical galaxies, such as the ones presented here. This stems
  from a combination of factors, but primarily from the sample
  selection. More massive and compact galaxies hosting brighter QSOs at
  stochastic peaks of their activity might be more affected by the
  growing SMBHs, while the more common activity linked to the lower
  luminosity AGN probed here does not appreciably affect the host. 
\end{enumerate}

Future observations, covering in particular CO transitions at high
  $J$, at high spatial resolution, and in large samples spanning a wide
  range of \fagn, will test our findings and address
  the issues on the coevolution of AGN and their hosts raised in this work.

\begin{acknowledgements}
  We acknowledge the constructive comments from the anonymous referee
  that greatly improved the content and presentation of the results.
  We warmly thank Marianne Vestergaard for providing a modified version of
  the \textsc{Linmix\_err.pro} routine for the linear regression with
  fixed slope and D. Elbaz and T. Wang for sharing their
  \textit{Herschel} catalog of GOODS-S.
  F.V. acknowledges support
  from the Carlsberg Foundation Research Grant CF18-0388
  ``Galaxies: Rise and Death''. F.V. and G.E.M. acknowledge the Villum
  Fonden Research Grant 13160 ``Gas to stars, stars to dust:
  tracing star formation across cosmic time'' and the Cosmic Dawn
  Center of Excellence funded by the Danish National Research
  Foundation under then Grant No. 140. A.P. gratefully acknowledges
  financial support from STFC through grants ST/T000244/1 and
  ST/P000541/1. Y.G.’s work is partially supported by National Key Basic
  Research and Development Program of China (grant
  No. 2017YFA0402704), National Natural Science Foundation of China
  (NSFC, Nos. 12033004, and 11861131007), and Chinese Academy of
  Sciences Key Research Program of Frontier Sciences (grant
  No. QYZDJ-SSW-SLH008). M.A. acknowledges support from FONDECYT grant
  1211951, “ANID+PCI+INSTITUTO MAX PLANCK DE ASTRONOMIA MPG 190030”
  and “ANID+PCI+REDES 190194. In this work,
  we made use of the COSMOS master spectroscopic catalog --
  kept updated by Mara Salvato --, of GILDAS, and STSDAS. GILDAS,
  the Grenoble Image and
  Line Data Analysis Software, is a joint effort of IRAM and the
  Observatoire de Grenoble. STSDAS is a product of the Space Telescope
  Science Institute, which is operated by AURA for NASA. Moreover,
  this paper makes use of the following ALMA data: ADS/JAO.ALMA,
  \#2019.1.01702.S, \#2018.1.00635.S, \#2016.1.01040.S, \#2016.1.00171.S, and
  \#2015.1.00260.S. ALMA is a partnership of ESO (representing its
  member states), NSF (USA) and NINS (Japan), together with NRC
  (Canada), MOST and ASIAA (Taiwan), and KASI (Republic of Korea),
  in cooperation with the Republic of Chile. The Joint ALMA
  Observatory is operated by ESO, AUI/NRAO, and NAOJ. 
\end{acknowledgements}

\bibliography{bib_agnexcitation} 

\begin{thebibliography}{132}
\expandafter\ifx\csname natexlab\endcsname\relax\def\natexlab#1{#1}\fi

\bibitem[{{Aird} {et~al.}(2015){Aird}, {Coil}, {Georgakakis}, {Nandra},
  {Barro}, \& {P{\'e}rez-Gonz{\'a}lez}}]{aird_2015}
{Aird}, J., {Coil}, A.~L., {Georgakakis}, A., {et~al.} 2015, \mnras, 451, 1892

\bibitem[{{Banerji} {et~al.}(2018){Banerji}, {Jones}, {Wagg}, {Carilli},
  {Bisbas}, \& {Hewett}}]{banerji_2018}
{Banerji}, M., {Jones}, G.~C., {Wagg}, J., {et~al.} 2018, \mnras, 479, 1154

\bibitem[{{Barnes} {et~al.}(2020){Barnes}, {Kannan}, {Vogelsberger}, \&
  {Marinacci}}]{barnes_2020}
{Barnes}, D.~J., {Kannan}, R., {Vogelsberger}, M., \& {Marinacci}, F. 2020,
  \mnras, 494, 1143

\bibitem[{{Bernhard} {et~al.}(2021){Bernhard}, {Tadhunter}, {Mullaney},
  {Grimmett}, {Rosario}, \& {Alexander}}]{bernhard_2021}
{Bernhard}, E., {Tadhunter}, C., {Mullaney}, J.~R., {et~al.} 2021, \mnras, 503,
  2598

\bibitem[{{Biernacki} \& {Teyssier}(2018)}]{biernacki_2018}
{Biernacki}, P. \& {Teyssier}, R. 2018, \mnras, 475, 5688

\bibitem[{{Bisbas} {et~al.}(2015){Bisbas}, {Papadopoulos}, \&
  {Viti}}]{bisbas_2015}
{Bisbas}, T.~G., {Papadopoulos}, P.~P., \& {Viti}, S. 2015, \apj, 803, 37

\bibitem[{{Bischetti} {et~al.}(2021){Bischetti}, {Feruglio}, {Piconcelli},
  {Duras}, {P{\'e}rez-Torres}, {Herrero}, {Venturi}, {Carniani}, {Bruni},
  {Gavignaud}, {Testa}, {Bongiorno}, {Brusa}, {Circosta}, {Cresci},
  {D'Odorico}, {Maiolino}, {Marconi}, {Mingozzi}, {Pappalardo}, {Perna},
  {Traianou}, {Travascio}, {Vietri}, {Zappacosta}, \& {Fiore}}]{bischetti_2021}
{Bischetti}, M., {Feruglio}, C., {Piconcelli}, E., {et~al.} 2021, \aap, 645,
  A33

\bibitem[{{Blain} {et~al.}(2003){Blain}, {Barnard}, \& {Chapman}}]{blain_2003}
{Blain}, A.~W., {Barnard}, V.~E., \& {Chapman}, S.~C. 2003, \mnras, 338, 733

\bibitem[{{Boogaard} {et~al.}(2020){Boogaard}, {van der Werf}, {Weiss},
  {Popping}, {Decarli}, {Walter}, {Aravena}, {Bouwens}, {Riechers},
  {Gonz{\'a}lez-L{\'o}pez}, {Smail}, {Carilli}, {Kaasinen}, {Daddi}, {Cox},
  {D{\'\i}az-Santos}, {Inami}, {Cortes}, \& {Wagg}}]{boogaard_2020}
{Boogaard}, L.~A., {van der Werf}, P., {Weiss}, A., {et~al.} 2020, \apj, 902,
  109

\bibitem[{{Bothwell} {et~al.}(2013){Bothwell}, {Smail}, {Chapman}, {Genzel},
  {Ivison}, {Tacconi}, {Alaghband-Zadeh}, {Bertoldi}, {Blain}, {Casey}, {Cox},
  {Greve}, {Lutz}, {Neri}, {Omont}, \& {Swinbank}}]{bothwell_2013}
{Bothwell}, M.~S., {Smail}, I., {Chapman}, S.~C., {et~al.} 2013, \mnras, 429,
  3047

\bibitem[{{Brown} {et~al.}(2019){Brown}, {Nayyeri}, {Cooray}, {Ma}, {Hickox},
  \& {Azadi}}]{brown_2019}
{Brown}, A., {Nayyeri}, H., {Cooray}, A., {et~al.} 2019, \apj, 871, 87

\bibitem[{{Brusa} {et~al.}(2018){Brusa}, {Cresci}, {Daddi}, {Paladino},
  {Perna}, {Bongiorno}, {Lusso}, {Sargent}, {Casasola}, {Feruglio},
  {Fraternali}, {Georgiev}, {Mainieri}, {Carniani}, {Comastri}, {Duras},
  {Fiore}, {Mannucci}, {Marconi}, {Piconcelli}, {Zamorani}, {Gilli}, {La
  Franca}, {Lanzuisi}, {Lutz}, {Santini}, {Scoville}, {Vignali}, {Vito},
  {Rabien}, {Busoni}, \& {Bonaglia}}]{brusa_2018}
{Brusa}, M., {Cresci}, G., {Daddi}, E., {et~al.} 2018, \aap, 612, A29

\bibitem[{{Carilli} \& {Walter}(2013)}]{carilli_2013}
{Carilli}, C.~L. \& {Walter}, F. 2013, \araa, 51, 105

\bibitem[{{Carniani} {et~al.}(2019){Carniani}, {Gallerani}, {Vallini},
  {Pallottini}, {Tazzari}, {Ferrara}, {Maiolino}, {Cicone}, {Feruglio}, {Neri},
  {D'Odorico}, {Wang}, \& {Li}}]{carniani_2019}
{Carniani}, S., {Gallerani}, S., {Vallini}, L., {et~al.} 2019, \mnras, 489,
  3939

\bibitem[{{Casey} {et~al.}(2014){Casey}, {Narayanan}, \& {Cooray}}]{casey_2014}
{Casey}, C.~M., {Narayanan}, D., \& {Cooray}, A. 2014, \physrep, 541, 45

\bibitem[{{Chabrier}(2003)}]{chabrier_2003}
{Chabrier}, G. 2003, \pasp, 115, 763

\bibitem[{{Cicone} {et~al.}(2018){Cicone}, {Brusa}, {Ramos Almeida}, {Cresci},
  {Husemann}, \& {Mainieri}}]{cicone_2018}
{Cicone}, C., {Brusa}, M., {Ramos Almeida}, C., {et~al.} 2018, Nature
  Astronomy, 2, 176

\bibitem[{{Cicone} {et~al.}(2012){Cicone}, {Feruglio}, {Maiolino}, {Fiore},
  {Piconcelli}, {Menci}, {Aussel}, \& {Sturm}}]{cicone_2012}
{Cicone}, C., {Feruglio}, C., {Maiolino}, R., {et~al.} 2012, \aap, 543, A99

\bibitem[{{Cicone} {et~al.}(2014){Cicone}, {Maiolino}, {Sturm},
  {Graci{\'a}-Carpio}, {Feruglio}, {Neri}, {Aalto}, {Davies}, {Fiore},
  {Fischer}, {Garc{\'\i}a-Burillo}, {Gonz{\'a}lez-Alfonso}, {Hailey-Dunsheath},
  {Piconcelli}, \& {Veilleux}}]{cicone_2014}
{Cicone}, C., {Maiolino}, R., {Sturm}, E., {et~al.} 2014, \aap, 562, A21

\bibitem[{{Ciesla} {et~al.}(2015){Ciesla}, {Charmandaris}, {Georgakakis},
  {Bernhard}, {Mitchell}, {Buat}, {Elbaz}, {LeFloc'h}, {Lacey}, {Magdis}, \&
  {Xilouris}}]{ciesla_2015}
{Ciesla}, L., {Charmandaris}, V., {Georgakakis}, A., {et~al.} 2015, \aap, 576,
  A10

\bibitem[{{Circosta} {et~al.}(2021){Circosta}, {Mainieri}, {Lamperti},
  {Padovani}, {Bischetti}, {Harrison}, {Kakkad}, {Zanella}, {Vietri},
  {Lanzuisi}, {Salvato}, {Brusa}, {Carniani}, {Cicone}, {Cresci}, {Feruglio},
  {Husemann}, {Mannucci}, {Marconi}, {Perna}, {Piconcelli}, {Puglisi},
  {Saintonge}, {Schramm}, {Vignali}, \& {Zappacosta}}]{circosta_2021}
{Circosta}, C., {Mainieri}, V., {Lamperti}, I., {et~al.} 2021, \aap, 646, A96

\bibitem[{{Circosta} {et~al.}(2018){Circosta}, {Mainieri}, {Padovani},
  {Lanzuisi}, {Salvato}, {Harrison}, {Kakkad}, {Puglisi}, {Vietri}, {Zamorani},
  {Cicone}, {Husemann}, {Vignali}, {Balmaverde}, {Bischetti}, {Bongiorno},
  {Brusa}, {Carniani}, {Civano}, {Comastri}, {Cresci}, {Feruglio}, {Fiore},
  {Fotopoulou}, {Karim}, {Lamastra}, {Magnelli}, {Mannucci}, {Marconi},
  {Merloni}, {Netzer}, {Perna}, {Piconcelli}, {Rodighiero}, {Schinnerer},
  {Schramm}, {Schulze}, {Silverman}, \& {Zappacosta}}]{circosta_2018}
{Circosta}, C., {Mainieri}, V., {Padovani}, P., {et~al.} 2018, \aap, 620, A82

\bibitem[{{Civano} {et~al.}(2016){Civano}, {Marchesi}, {Comastri}, {Urry},
  {Elvis}, {Cappelluti}, {Puccetti}, {Brusa}, {Zamorani}, {Hasinger},
  {Aldcroft}, {Alexander}, {Allevato}, {Brunner}, {Capak}, {Finoguenov},
  {Fiore}, {Fruscione}, {Gilli}, {Glotfelty}, {Griffiths}, {Hao}, {Harrison},
  {Jahnke}, {Kartaltepe}, {Karim}, {LaMassa}, {Lanzuisi}, {Miyaji}, {Ranalli},
  {Salvato}, {Sargent}, {Scoville}, {Schawinski}, {Schinnerer}, {Silverman},
  {Smolcic}, {Stern}, {Toft}, {Trakhenbrot}, {Treister}, \&
  {Vignali}}]{civano_2016}
{Civano}, F., {Marchesi}, S., {Comastri}, A., {et~al.} 2016, \apj, 819, 62

\bibitem[{{Cortzen} {et~al.}(2020){Cortzen}, {Magdis}, {Valentino}, {Daddi},
  {Liu}, {Rigopoulou}, {Sargent}, {Riechers}, {Cormier}, {Hodge}, {Walter},
  {Elbaz}, {B{\'e}thermin}, {Greve}, {Kokorev}, \& {Toft}}]{cortzen_2020}
{Cortzen}, I., {Magdis}, G.~E., {Valentino}, F., {et~al.} 2020, \aap, 634, L14

\bibitem[{{Costa} {et~al.}(2020){Costa}, {Pakmor}, \& {Springel}}]{costa_2020}
{Costa}, T., {Pakmor}, R., \& {Springel}, V. 2020, \mnras, 497, 5229

\bibitem[{{Croton} {et~al.}(2006){Croton}, {Springel}, {White}, {De Lucia},
  {Frenk}, {Gao}, {Jenkins}, {Kauffmann}, {Navarro}, \&
  {Yoshida}}]{croton_2006}
{Croton}, D.~J., {Springel}, V., {White}, S. D.~M., {et~al.} 2006, \mnras, 365,
  11

\bibitem[{{Daddi} {et~al.}(2015){Daddi}, {Dannerbauer}, {Liu}, {Aravena},
  {Bournaud}, {Walter}, {Riechers}, {Magdis}, {Sargent}, {B{\'e}thermin},
  {Carilli}, {Cibinel}, {Dickinson}, {Elbaz}, {Gao}, {Gobat}, {Hodge}, \&
  {Krips}}]{daddi_2015}
{Daddi}, E., {Dannerbauer}, H., {Liu}, D., {et~al.} 2015, \aap, 577, A46

\bibitem[{{D'Amato} {et~al.}(2020){D'Amato}, {Gilli}, {Vignali}, {Massardi},
  {Pozzi}, {Zamorani}, {Circosta}, {Vito}, {Fritz}, {Cresci}, {Casasola},
  {Calura}, {Feltre}, {Manieri}, {Rigopoulou}, {Tozzi}, \&
  {Norman}}]{damato_2020}
{D'Amato}, Q., {Gilli}, R., {Vignali}, C., {et~al.} 2020, \aap, 636, A37

\bibitem[{{Del Moro} {et~al.}(2016){Del Moro}, {Alexander}, {Bauer}, {Daddi},
  {Kocevski}, {McIntosh}, {Stanley}, {Brandt}, {Elbaz}, {Harrison}, {Luo},
  {Mullaney}, \& {Xue}}]{delmoro_2016}
{Del Moro}, A., {Alexander}, D.~M., {Bauer}, F.~E., {et~al.} 2016, \mnras, 456,
  2105

\bibitem[{{Delvecchio} {et~al.}(2020){Delvecchio}, {Daddi}, {Aird}, {Mullaney},
  {Bernhard}, {Grimmett}, {Carraro}, {Cimatti}, {Zamorani}, {Caplar}, {Vito},
  {Elbaz}, \& {Rodighiero}}]{delvecchio_2020}
{Delvecchio}, I., {Daddi}, E., {Aird}, J., {et~al.} 2020, \apj, 892, 17

\bibitem[{{Donley} {et~al.}(2012){Donley}, {Koekemoer}, {Brusa}, {Capak},
  {Cardamone}, {Civano}, {Ilbert}, {Impey}, {Kartaltepe}, {Miyaji}, {Salvato},
  {Sanders}, {Trump}, \& {Zamorani}}]{donley_2012}
{Donley}, J.~L., {Koekemoer}, A.~M., {Brusa}, M., {et~al.} 2012, \apj, 748, 142

\bibitem[{{Draine} \& {Li}(2007)}]{draine_2007}
{Draine}, B.~T. \& {Li}, A. 2007, \apj, 657, 810

\bibitem[{{Duras} {et~al.}(2020){Duras}, {Bongiorno}, {Ricci}, {Piconcelli},
  {Shankar}, {Lusso}, {Bianchi}, {Fiore}, {Maiolino}, {Marconi}, {Onori},
  {Sani}, {Schneider}, {Vignali}, \& {La Franca}}]{duras_2020}
{Duras}, F., {Bongiorno}, A., {Ricci}, F., {et~al.} 2020, \aap, 636, A73

\bibitem[{{Elbaz} {et~al.}(2018){Elbaz}, {Leiton}, {Nagar}, {Okumura},
  {Franco}, {Schreiber}, {Pannella}, {Wang}, {Dickinson}, {D{\'\i}az-Santos},
  {Ciesla}, {Daddi}, {Bournaud}, {Magdis}, {Zhou}, \&
  {Rujopakarn}}]{elbaz_2018}
{Elbaz}, D., {Leiton}, R., {Nagar}, N., {et~al.} 2018, \aap, 616, A110

\bibitem[{{Elvis} {et~al.}(1994){Elvis}, {Wilkes}, {McDowell}, {Green},
  {Bechtold}, {Willner}, {Oey}, {Polomski}, \& {Cutri}}]{elvis_1994}
{Elvis}, M., {Wilkes}, B.~J., {McDowell}, J.~C., {et~al.} 1994, \apjs, 95, 1

\bibitem[{{Feruglio} {et~al.}(2015){Feruglio}, {Fiore}, {Carniani},
  {Piconcelli}, {Zappacosta}, {Bongiorno}, {Cicone}, {Maiolino}, {Marconi},
  {Menci}, {Puccetti}, \& {Veilleux}}]{feruglio_2015}
{Feruglio}, C., {Fiore}, F., {Carniani}, S., {et~al.} 2015, \aap, 583, A99

\bibitem[{{Fogasy} {et~al.}(2020){Fogasy}, {Knudsen}, {Drouart}, {Lagos}, \&
  {Fan}}]{fogasy_2020}
{Fogasy}, J., {Knudsen}, K.~K., {Drouart}, G., {Lagos}, C.~D.~P., \& {Fan}, L.
  2020, \mnras, 493, 3744

\bibitem[{{F{\"o}rster Schreiber} {et~al.}(2014){F{\"o}rster Schreiber},
  {Genzel}, {Newman}, {Kurk}, {Lutz}, {Tacconi}, {Wuyts}, {Bandara}, {Burkert},
  {Buschkamp}, {Carollo}, {Cresci}, {Daddi}, {Davies}, {Eisenhauer}, {Hicks},
  {Lang}, {Lilly}, {Mainieri}, {Mancini}, {Naab}, {Peng}, {Renzini}, {Rosario},
  {Shapiro Griffin}, {Shapley}, {Sternberg}, {Tacchella}, {Vergani},
  {Wisnioski}, {Wuyts}, \& {Zamorani}}]{forster-schreiber_2014}
{F{\"o}rster Schreiber}, N.~M., {Genzel}, R., {Newman}, S.~F., {et~al.} 2014,
  \apj, 787, 38

\bibitem[{{Gabor} \& {Bournaud}(2014)}]{gabor_2014}
{Gabor}, J.~M. \& {Bournaud}, F. 2014, \mnras, 437, L56

\bibitem[{{Gallerani} {et~al.}(2014){Gallerani}, {Ferrara}, {Neri}, \&
  {Maiolino}}]{gallerani_2014}
{Gallerani}, S., {Ferrara}, A., {Neri}, R., \& {Maiolino}, R. 2014, \mnras,
  445, 2848

\bibitem[{{Gandhi} {et~al.}(2009){Gandhi}, {Horst}, {Smette}, {H{\"o}nig},
  {Comastri}, {Gilli}, {Vignali}, \& {Duschl}}]{gandhi_2009}
{Gandhi}, P., {Horst}, H., {Smette}, A., {et~al.} 2009, \aap, 502, 457

\bibitem[{{Greve} {et~al.}(2014){Greve}, {Leonidaki}, {Xilouris}, {Wei{\ss}},
  {Zhang}, {van der Werf}, {Aalto}, {Armus}, {D{\'\i}az-Santos}, {Evans},
  {Fischer}, {Gao}, {Gonz{\'a}lez-Alfonso}, {Harris}, {Henkel}, {Meijerink},
  {Naylor}, {Smith}, {Spaans}, {Stacey}, {Veilleux}, \& {Walter}}]{greve_2014}
{Greve}, T.~R., {Leonidaki}, I., {Xilouris}, E.~M., {et~al.} 2014, \apj, 794,
  142

\bibitem[{{Harrington} {et~al.}(2021){Harrington}, {Weiss}, {Yun}, {Magnelli},
  {Sharon}, {Leung}, {Vishwas}, {Wang}, {Frayer}, {Jim{\'e}nez-Andrade}, {Liu},
  {Garc{\'\i}a}, {Romano-D{\'\i}az}, {Frye}, {Jarugula}, {B{\u{a}}descu},
  {Berman}, {Dannerbauer}, {D{\'\i}az-S{\'a}nchez}, {Grassitelli},
  {Kamieneski}, {Kim}, {Kirkpatrick}, {Lowenthal}, {Messias}, {Puschnig},
  {Stacey}, {Torne}, \& {Bertoldi}}]{harrington_2021}
{Harrington}, K.~C., {Weiss}, A., {Yun}, M.~S., {et~al.} 2021, \apj, 908, 95

\bibitem[{{Harrison}(2017)}]{harrison_2017}
{Harrison}, C.~M. 2017, Nature Astronomy, 1, 0165

\bibitem[{{Harrison} {et~al.}(2015){Harrison}, {Thomson}, {Alexander}, {Bauer},
  {Edge}, {Hogan}, {Mullaney}, \& {Swinbank}}]{harrison_2015}
{Harrison}, C.~M., {Thomson}, A.~P., {Alexander}, D.~M., {et~al.} 2015, \apj,
  800, 45

\bibitem[{{Hickox} {et~al.}(2014){Hickox}, {Mullaney}, {Alexander}, {Chen},
  {Civano}, {Goulding}, \& {Hainline}}]{hickox_2014}
{Hickox}, R.~C., {Mullaney}, J.~R., {Alexander}, D.~M., {et~al.} 2014, \apj,
  782, 9

\bibitem[{{Hopkins} {et~al.}(2006){Hopkins}, {Hernquist}, {Cox}, {Di Matteo},
  {Robertson}, \& {Springel}}]{hopkins_2006}
{Hopkins}, P.~F., {Hernquist}, L., {Cox}, T.~J., {et~al.} 2006, \apjs, 163, 1

\bibitem[{{Israel} \& {Baas}(2002)}]{israel_2002}
{Israel}, F.~P. \& {Baas}, F. 2002, \aap, 383, 82

\bibitem[{{Israel} {et~al.}(2015){Israel}, {Rosenberg}, \& {van der
  Werf}}]{israel_2015}
{Israel}, F.~P., {Rosenberg}, M.~J.~F., \& {van der Werf}, P. 2015, \aap, 578,
  A95

\bibitem[{{Izumi} {et~al.}(2020){Izumi}, {Nguyen}, {Imanishi}, {Kawamuro},
  {Baba}, {Nakano}, {Kohno}, {Matsushita}, {Meier}, {Turner}, {Michiyama},
  {Harada}, {Mart{\'\i}n}, {Nakanishi}, {Takano}, {Wiklind}, {Nakai}, \&
  {Hsieh}}]{izumi_2020}
{Izumi}, T., {Nguyen}, D.~D., {Imanishi}, M., {et~al.} 2020, \apj, 898, 75

\bibitem[{{Jarvis} {et~al.}(2020){Jarvis}, {Harrison}, {Mainieri}, {Calistro
  Rivera}, {Jethwa}, {Zhang}, {Alexander}, {Circosta}, {Costa}, {De Breuck},
  {Kakkad}, {Kharb}, {Lansbury}, \& {Thomson}}]{jarvis_2020}
{Jarvis}, M.~E., {Harrison}, C.~M., {Mainieri}, V., {et~al.} 2020, \mnras, 498,
  1560

\bibitem[{{Jarvis} {et~al.}(2019){Jarvis}, {Harrison}, {Thomson}, {Circosta},
  {Mainieri}, {Alexander}, {Edge}, {Lansbury}, {Molyneux}, \&
  {Mullaney}}]{jarvis_2019}
{Jarvis}, M.~E., {Harrison}, C.~M., {Thomson}, A.~P., {et~al.} 2019, \mnras,
  485, 2710

\bibitem[{{Jiao} {et~al.}(2019){Jiao}, {Zhao}, {Lu}, {Gao}, {Salak}, {Zhu},
  {Zhang}, {Jiang}, \& {Tan}}]{jiao_2019}
{Jiao}, Q., {Zhao}, Y., {Lu}, N., {et~al.} 2019, \apj, 880, 133

\bibitem[{{Jiao} {et~al.}(2017){Jiao}, {Zhao}, {Zhu}, {Lu}, {Gao}, \&
  {Zhang}}]{jiao_2017}
{Jiao}, Q., {Zhao}, Y., {Zhu}, M., {et~al.} 2017, \apjl, 840, L18

\bibitem[{{Jin} {et~al.}(2018){Jin}, {Daddi}, {Liu}, {Smol{\v c}i{\'c}},
  {Schinnerer}, {Calabr{\`o}}, {Gu}, {Delhaize}, {Delvecchio}, {Gao},
  {Salvato}, {Puglisi}, {Dickinson}, {Bertoldi}, {Sargent}, {Novak}, {Magdis},
  {Aretxaga}, {Wilson}, \& {Capak}}]{jin_2018}
{Jin}, S., {Daddi}, E., {Liu}, D., {et~al.} 2018, \apj, 864, 56

\bibitem[{{Jin} {et~al.}(2019){Jin}, {Daddi}, {Magdis}, {Liu}, {Schinnerer},
  {Papadopoulos}, {Gu}, {Gao}, \& {Calabr{\`o}}}]{jin_2019}
{Jin}, S., {Daddi}, E., {Magdis}, G.~E., {et~al.} 2019, \apj, 887, 144

\bibitem[{{Kakkad} {et~al.}(2017){Kakkad}, {Mainieri}, {Brusa}, {Padovani},
  {Carniani}, {Feruglio}, {Sargent}, {Husemann}, {Bongiorno}, {Bonzini},
  {Piconcelli}, {Silverman}, \& {Rujopakarn}}]{kakkad_2017}
{Kakkad}, D., {Mainieri}, V., {Brusa}, M., {et~al.} 2017, \mnras, 468, 4205

\bibitem[{{Kamenetzky} {et~al.}(2016){Kamenetzky}, {Rangwala}, {Glenn},
  {Maloney}, \& {Conley}}]{kamenetzky_2016}
{Kamenetzky}, J., {Rangwala}, N., {Glenn}, J., {Maloney}, P.~R., \& {Conley},
  A. 2016, \apj, 829, 93

\bibitem[{Kaplan \& Meier(1958)}]{kaplan-meier_1958}
Kaplan, E.~L. \& Meier, P. 1958, J. Am. Stat. Assoc., 53, 457

\bibitem[{{Kelly}(2007)}]{kelly_2007}
{Kelly}, B.~C. 2007, \apj, 665, 1489

\bibitem[{{Kilerci Eser} {et~al.}(2015){Kilerci Eser}, {Vestergaard},
  {Peterson}, {Denney}, \& {Bentz}}]{kilerci-eser_2015}
{Kilerci Eser}, E., {Vestergaard}, M., {Peterson}, B.~M., {Denney}, K.~D., \&
  {Bentz}, M.~C. 2015, \apj, 801, 8

\bibitem[{{Kirkpatrick} {et~al.}(2017){Kirkpatrick}, {Alberts}, {Pope},
  {Barro}, {Bonato}, {Kocevski}, {P{\'e}rez-Gonz{\'a}lez}, {Rieke},
  {Rodr{\'\i}guez-Mu{\~n}oz}, {Sajina}, {Grogin}, {Mantha}, {Pandya}, {Pforr},
  {Salvato}, \& {Santini}}]{kirkpatrick_2017}
{Kirkpatrick}, A., {Alberts}, S., {Pope}, A., {et~al.} 2017, \apj, 849, 111

\bibitem[{{Kirkpatrick} {et~al.}(2019){Kirkpatrick}, {Sharon}, {Keller}, \&
  {Pope}}]{kirkpatrick_2019}
{Kirkpatrick}, A., {Sharon}, C., {Keller}, E., \& {Pope}, A. 2019, \apj, 879,
  41

\bibitem[{{Kokorev} {et~al.}(2021){Kokorev}, {Magdis}, \&
  {Davidzon}}]{kokorev_2021}
{Kokorev}, V.~I., {Magdis}, G.~E., \& {Davidzon}, I. 2021, \apj, in press

\bibitem[{{Laigle} {et~al.}(2016){Laigle}, {McCracken}, {Ilbert}, {Hsieh},
  {Davidzon}, {Capak}, {Hasinger}, {Silverman}, {Pichon}, {Coupon}, {Aussel},
  {Le Borgne}, {Caputi}, {Cassata}, {Chang}, {Civano}, {Dunlop}, {Fynbo},
  {Kartaltepe}, {Koekemoer}, {Le F{\`e}vre}, {Le Floc'h}, {Leauthaud}, {Lilly},
  {Lin}, {Marchesi}, {Milvang-Jensen}, {Salvato}, {Sanders}, {Scoville},
  {Smolcic}, {Stockmann}, {Taniguchi}, {Tasca}, {Toft}, {Vaccari}, \&
  {Zabl}}]{laigle_2016}
{Laigle}, C., {McCracken}, H.~J., {Ilbert}, O., {et~al.} 2016, \apjs, 224, 24

\bibitem[{{Lehmer} {et~al.}(2010){Lehmer}, {Alexander}, {Bauer}, {Brandt},
  {Goulding}, {Jenkins}, {Ptak}, \& {Roberts}}]{lehmer_2010}
{Lehmer}, B.~D., {Alexander}, D.~M., {Bauer}, F.~E., {et~al.} 2010, \apj, 724,
  559

\bibitem[{{Li} \& {Draine}(2001)}]{li_2001}
{Li}, A. \& {Draine}, B.~T. 2001, \apj, 554, 778

\bibitem[{{Liu} {et~al.}(2021){Liu}, {Daddi}, {Schinnerer}, {Saito}, {Leroy},
  {Silverman}, {Valentino}, {Magdis}, {Gao}, {Jin}, {Puglisi}, \&
  {Groves}}]{liu_2021}
{Liu}, D., {Daddi}, E., {Schinnerer}, E., {et~al.} 2021, \apj, 909, 56

\bibitem[{{Liu} {et~al.}(2015){Liu}, {Gao}, {Isaak}, {Daddi}, {Yang}, {Lu}, \&
  {van der Werf}}]{liu_2015}
{Liu}, D., {Gao}, Y., {Isaak}, K., {et~al.} 2015, \apjl, 810, L14

\bibitem[{{Lu} {et~al.}(2017){Lu}, {Zhao}, {D{\'{\i}}az-Santos}, {Xu}, {Gao},
  {Armus}, {Isaak}, {Mazzarella}, {van der Werf}, {Appleton}, {Charmandaris},
  {Evans}, {Howell}, {Iwasawa}, {Leech}, {Lord}, {Petric}, {Privon}, {Sanders},
  {Schulz}, \& {Surace}}]{lu_2017}
{Lu}, N., {Zhao}, Y., {D{\'{\i}}az-Santos}, T., {et~al.} 2017, \apjs, 230, 1

\bibitem[{{Lu} {et~al.}(2014){Lu}, {Zhao}, {Xu}, {Gao}, {Armus}, {Mazzarella},
  {Isaak}, {Petric}, {Charmand aris}, {D{\'\i}az-Santos}, {Evans}, {Howell},
  {Appleton}, {Inami}, {Iwasawa}, {Leech}, {Lord}, {Sanders}, {Schulz},
  {Surace}, \& {van der Werf}}]{lu_2014}
{Lu}, N., {Zhao}, Y., {Xu}, C.~K., {et~al.} 2014, \apjl, 787, L23

\bibitem[{{Lu} {et~al.}(2015){Lu}, {Zhao}, {Xu}, {Gao}, {D{\'\i}az-Santos},
  {Charmandaris}, {Inami}, {Howell}, {Liu}, {Armus}, {Mazzarella}, {Privon},
  {Lord}, {Sanders}, {Schulz}, \& {van der Werf}}]{lu_2015}
{Lu}, N., {Zhao}, Y., {Xu}, C.~K., {et~al.} 2015, \apjl, 802, L11

\bibitem[{{Lusso} {et~al.}(2012){Lusso}, {Comastri}, {Simmons}, {Mignoli},
  {Zamorani}, {Vignali}, {Brusa}, {Shankar}, {Lutz}, {Trump}, {Maiolino},
  {Gilli}, {Bolzonella}, {Puccetti}, {Salvato}, {Impey}, {Civano}, {Elvis},
  {Mainieri}, {Silverman}, {Koekemoer}, {Bongiorno}, {Merloni}, {Berta}, {Le
  Floc'h}, {Magnelli}, {Pozzi}, \& {Riguccini}}]{lusso_2012}
{Lusso}, E., {Comastri}, A., {Simmons}, B.~D., {et~al.} 2012, \mnras, 425, 623

\bibitem[{{Lusso} {et~al.}(2011){Lusso}, {Comastri}, {Vignali}, {Zamorani},
  {Treister}, {Sanders}, {Bolzonella}, {Bongiorno}, {Brusa}, {Civano}, {Gilli},
  {Mainieri}, {Nair}, {Aller}, {Carollo}, {Koekemoer}, {Merloni}, \&
  {Trump}}]{lusso_2011}
{Lusso}, E., {Comastri}, A., {Vignali}, C., {et~al.} 2011, \aap, 534, A110

\bibitem[{{Lutz} {et~al.}(2011){Lutz}, {Poglitsch}, {Altieri}, {Andreani},
  {Aussel}, {Berta}, {Bongiovanni}, {Brisbin}, {Cava}, {Cepa}, {Cimatti},
  {Daddi}, {Dominguez-Sanchez}, {Elbaz}, {F{\"o}rster Schreiber}, {Genzel},
  {Grazian}, {Gruppioni}, {Harwit}, {Le Floc'h}, {Magdis}, {Magnelli},
  {Maiolino}, {Nordon}, {P{\'e}rez Garc{\'{\i}}a}, {Popesso}, {Pozzi},
  {Riguccini}, {Rodighiero}, {Saintonge}, {Sanchez Portal}, {Santini}, {Shao},
  {Sturm}, {Tacconi}, {Valtchanov}, {Wetzstein}, \& {Wieprecht}}]{lutz_2011}
{Lutz}, D., {Poglitsch}, A., {Altieri}, B., {et~al.} 2011, \aap, 532, A90

\bibitem[{{Madden} {et~al.}(2020){Madden}, {Cormier}, {Hony}, {Lebouteiller},
  {Abel}, {Galametz}, {De Looze}, {Chevance}, {Polles}, {Lee}, {Galliano},
  {Lambert-Huyghe}, {Hu}, \& {Ramambason}}]{madden_2020}
{Madden}, S.~C., {Cormier}, D., {Hony}, S., {et~al.} 2020, \aap, 643, A141

\bibitem[{{Magdis} {et~al.}(2012){Magdis}, {Daddi}, {B{\'e}thermin}, {Sargent},
  {Elbaz}, {Pannella}, {Dickinson}, {Dannerbauer}, {da Cunha}, {Walter},
  {Rigopoulou}, {Charmandaris}, {Hwang}, \& {Kartaltepe}}]{magdis_2012}
{Magdis}, G.~E., {Daddi}, E., {B{\'e}thermin}, M., {et~al.} 2012, \apj, 760, 6

\bibitem[{{Magdis} {et~al.}(2017){Magdis}, {Rigopoulou}, {Daddi}, {Bethermin},
  {Feruglio}, {Sargent}, {Dannerbauer}, {Dickinson}, {Elbaz}, {Gomez Guijarro},
  {Huang}, {Toft}, \& {Valentino}}]{magdis_2017}
{Magdis}, G.~E., {Rigopoulou}, D., {Daddi}, E., {et~al.} 2017, \aap, 603, A93

\bibitem[{{Marchesi} {et~al.}(2016){Marchesi}, {Civano}, {Elvis}, {Salvato},
  {Brusa}, {Comastri}, {Gilli}, {Hasinger}, {Lanzuisi}, {Miyaji}, {Treister},
  {Urry}, {Vignali}, {Zamorani}, {Allevato}, {Cappelluti}, {Cardamone},
  {Finoguenov}, {Griffiths}, {Karim}, {Laigle}, {LaMassa}, {Jahnke}, {Ranalli},
  {Schawinski}, {Schinnerer}, {Silverman}, {Smolcic}, {Suh}, \&
  {Trakhtenbrot}}]{marchesi_2016}
{Marchesi}, S., {Civano}, F., {Elvis}, M., {et~al.} 2016, \apj, 817, 34

\bibitem[{{Mashian} {et~al.}(2015){Mashian}, {Sturm}, {Sternberg}, {Janssen},
  {Hailey-Dunsheath}, {Fischer}, {Contursi}, {Gonz{\'a}lez-Alfonso},
  {Graci{\'a}-Carpio}, {Poglitsch}, {Veilleux}, {Davies}, {Genzel}, {Lutz},
  {Tacconi}, {Verma}, {Wei{\ss}}, {Polisensky}, \& {Nikola}}]{mashian_2015}
{Mashian}, N., {Sturm}, E., {Sternberg}, A., {et~al.} 2015, \apj, 802, 81

\bibitem[{{Meijerink} {et~al.}(2013){Meijerink}, {Kristensen}, {Wei{\ss}}, {van
  der Werf}, {Walter}, {Spaans}, {Loenen}, {Fischer}, {Israel}, {Isaak},
  {Papadopoulos}, {Aalto}, {Armus}, {Charmandaris}, {Dasyra}, {Diaz-Santos},
  {Evans}, {Gao}, {Gonz{\'a}lez-Alfonso}, {G{\"u}sten}, {Henkel}, {Kramer},
  {Lord}, {Mart{\'\i}n-Pintado}, {Naylor}, {Sanders}, {Smith}, {Spinoglio},
  {Stacey}, {Veilleux}, \& {Wiedner}}]{meijerink_2013}
{Meijerink}, R., {Kristensen}, L.~E., {Wei{\ss}}, A., {et~al.} 2013, \apjl,
  762, L16

\bibitem[{{Meijerink} {et~al.}(2007){Meijerink}, {Spaans}, \&
  {Israel}}]{meijerink_2007}
{Meijerink}, R., {Spaans}, M., \& {Israel}, F.~P. 2007, \aap, 461, 793

\bibitem[{{Mullaney} {et~al.}(2011){Mullaney}, {Alexander}, {Goulding}, \&
  {Hickox}}]{mullaney_2011}
{Mullaney}, J.~R., {Alexander}, D.~M., {Goulding}, A.~D., \& {Hickox}, R.~C.
  2011, \mnras, 414, 1082

\bibitem[{{Mullaney} {et~al.}(2012){Mullaney}, {Pannella}, {Daddi},
  {Alexander}, {Elbaz}, {Hickox}, {Bournaud}, {Altieri}, {Aussel}, {Coia},
  {Dannerbauer}, {Dasyra}, {Dickinson}, {Hwang}, {Kartaltepe}, {Leiton},
  {Magdis}, {Magnelli}, {Popesso}, {Valtchanov}, {Bauer}, {Brandt}, {Del Moro},
  {Hanish}, {Ivison}, {Juneau}, {Luo}, {Lutz}, {Sargent}, {Scott}, \&
  {Xue}}]{mullaney_2012}
{Mullaney}, J.~R., {Pannella}, M., {Daddi}, E., {et~al.} 2012, \mnras, 419, 95

\bibitem[{{Muzzin} {et~al.}(2013){Muzzin}, {Wilson}, {Demarco}, {Lidman},
  {Nantais}, {Hoekstra}, {Yee}, \& {Rettura}}]{muzzin_2013}
{Muzzin}, A., {Wilson}, G., {Demarco}, R., {et~al.} 2013, \apj, 767, 39

\bibitem[{{Narayanan} \& {Krumholz}(2014)}]{narayanan_2014}
{Narayanan}, D. \& {Krumholz}, M.~R. 2014, \mnras, 442, 1411

\bibitem[{{Noll} {et~al.}(2009){Noll}, {Pierini}, {Cimatti}, {Daddi}, {Kurk},
  {Bolzonella}, {Cassata}, {Halliday}, {Mignoli}, {Pozzetti}, {Renzini},
  {Berta}, {Dickinson}, {Franceschini}, {Rodighiero}, {Rosati}, \&
  {Zamorani}}]{noll_2009}
{Noll}, S., {Pierini}, D., {Cimatti}, A., {et~al.} 2009, \aap, 499, 69

\bibitem[{{Papadopoulos} {et~al.}(2018){Papadopoulos}, {Bisbas}, \&
  {Zhang}}]{papadopoulos_2018}
{Papadopoulos}, P.~P., {Bisbas}, T.~G., \& {Zhang}, Z.-Y. 2018, \mnras, 478,
  1716

\bibitem[{{Papadopoulos} {et~al.}(2004){Papadopoulos}, {Thi}, \&
  {Viti}}]{papadopoulos_2004}
{Papadopoulos}, P.~P., {Thi}, W.-F., \& {Viti}, S. 2004, \mnras, 351, 147

\bibitem[{{Perna} {et~al.}(2015){Perna}, {Brusa}, {Salvato}, {Cresci},
  {Lanzuisi}, {Berta}, {Delvecchio}, {Fiore}, {Lutz}, {Le Floc'h}, {Mainieri},
  \& {Riguccini}}]{perna_2015}
{Perna}, M., {Brusa}, M., {Salvato}, M., {et~al.} 2015, \aap, 583, A72

\bibitem[{{Perna} {et~al.}(2018){Perna}, {Sargent}, {Brusa}, {Daddi},
  {Feruglio}, {Cresci}, {Lanzuisi}, {Lusso}, {Comastri}, {Coogan}, {D'Amato},
  {Gilli}, {Piconcelli}, \& {Vignali}}]{perna_2018}
{Perna}, M., {Sargent}, M.~T., {Brusa}, M., {et~al.} 2018, \aap, 619, A90

\bibitem[{{Puglisi} {et~al.}(2021{\natexlab{a}}){Puglisi}, {Daddi}, {Brusa},
  {Bournaud}, {Fensch}, {Liu}, {Delvecchio}, {Calabr{\`o}}, {Circosta},
  {Valentino}, {Perna}, {Jin}, {Enia}, {Mancini}, \&
  {Rodighiero}}]{puglisi_2021nat}
{Puglisi}, A., {Daddi}, E., {Brusa}, M., {et~al.} 2021{\natexlab{a}}, Nature
  Astronomy, 5, 319

\bibitem[{{Puglisi} {et~al.}(2019){Puglisi}, {Daddi}, {Liu}, {Bournaud},
  {Silverman}, {Circosta}, {Calabr{\`o}}, {Aravena}, {Cibinel}, \&
  {Dannerbauer}}]{puglisi_2019}
{Puglisi}, A., {Daddi}, E., {Liu}, D., {et~al.} 2019, \apj, 877, L23

\bibitem[{{Puglisi} {et~al.}(2021{\natexlab{b}}){Puglisi}, {Daddi},
  {Valentino}, {Magdis}, {Liu}, {Kokorev}, {Circosta}, {Elbaz}, {Bournaud},
  {Gomez-Guijarro}, {Jin}, {Madden}, {Sargent}, \& {Swinbank}}]{puglisi_2021}
{Puglisi}, A., {Daddi}, E., {Valentino}, F., {et~al.} 2021{\natexlab{b}}, arXiv
  e-prints, arXiv:2103.12035

\bibitem[{{Ranalli} {et~al.}(2003){Ranalli}, {Comastri}, \&
  {Setti}}]{ranalli_2003}
{Ranalli}, P., {Comastri}, A., \& {Setti}, G. 2003, \aap, 399, 39

\bibitem[{{Rodighiero} {et~al.}(2011){Rodighiero}, {Daddi}, {Baronchelli},
  {Cimatti}, {Renzini}, {Aussel}, {Popesso}, {Lutz}, {Andreani}, {Berta},
  {Cava}, {Elbaz}, {Feltre}, {Fontana}, {F{\"o}rster Schreiber},
  {Franceschini}, {Genzel}, {Grazian}, {Gruppioni}, {Ilbert}, {Le Floch},
  {Magdis}, {Magliocchetti}, {Magnelli}, {Maiolino}, {McCracken}, {Nordon},
  {Poglitsch}, {Santini}, {Pozzi}, {Riguccini}, {Tacconi}, {Wuyts}, \&
  {Zamorani}}]{rodighiero_2011}
{Rodighiero}, G., {Daddi}, E., {Baronchelli}, I., {et~al.} 2011, \apjl, 739,
  L40

\bibitem[{{Roos} {et~al.}(2015){Roos}, {Juneau}, {Bournaud}, \&
  {Gabor}}]{roos_2015}
{Roos}, O., {Juneau}, S., {Bournaud}, F., \& {Gabor}, J.~M. 2015, \apj, 800, 19

\bibitem[{{Rosenberg} {et~al.}(2015){Rosenberg}, {van der Werf}, {Aalto},
  {Armus}, {Charmandaris}, {D{\'\i}az-Santos}, {Evans}, {Fischer}, {Gao},
  {Gonz{\'a}lez-Alfonso}, {Greve}, {Harris}, {Henkel}, {Israel}, {Isaak},
  {Kramer}, {Meijerink}, {Naylor}, {Sanders}, {Smith}, {Spaans}, {Spinoglio},
  {Stacey}, {Veenendaal}, {Veilleux}, {Walter}, {Wei{\ss}}, {Wiedner}, {van der
  Wiel}, \& {Xilouris}}]{rosenberg_2015}
{Rosenberg}, M.~J.~F., {van der Werf}, P.~P., {Aalto}, S., {et~al.} 2015, \apj,
  801, 72

\bibitem[{{Saintonge} {et~al.}(2012){Saintonge}, {Tacconi}, {Fabello}, {Wang},
  {Catinella}, {Genzel}, {Graci{\'a}-Carpio}, {Kramer}, {Moran}, {Heckman},
  {Schiminovich}, {Schuster}, \& {Wuyts}}]{saintonge_2012}
{Saintonge}, A., {Tacconi}, L.~J., {Fabello}, S., {et~al.} 2012, \apj, 758, 73

\bibitem[{{Sargent} {et~al.}(2014){Sargent}, {Daddi}, {B{\'e}thermin},
  {Aussel}, {Magdis}, {Hwang}, {Juneau}, {Elbaz}, \& {da Cunha}}]{sargent_2014}
{Sargent}, M.~T., {Daddi}, E., {B{\'e}thermin}, M., {et~al.} 2014, \apj, 793,
  19

\bibitem[{{Schulze} {et~al.}(2019){Schulze}, {Silverman}, {Daddi},
  {Rujopakarn}, {Liu}, {Schramm}, {Mainieri}, {Imanishi}, {Hirschmann}, \&
  {Jahnke}}]{schulze_2019}
{Schulze}, A., {Silverman}, J.~D., {Daddi}, E., {et~al.} 2019, \mnras, 488,
  1180

\bibitem[{{Scoville} {et~al.}(2007){Scoville}, {Aussel}, {Brusa}, {Capak},
  {Carollo}, {Elvis}, {Giavalisco}, {Guzzo}, {Hasinger}, {Impey}, {Kneib},
  {LeFevre}, {Lilly}, {Mobasher}, {Renzini}, {Rich}, {Sanders}, {Schinnerer},
  {Schminovich}, {Shopbell}, {Taniguchi}, \& {Tyson}}]{scoville_2007}
{Scoville}, N., {Aussel}, H., {Brusa}, M., {et~al.} 2007, \apjs, 172, 1

\bibitem[{{Scoville} {et~al.}(2017){Scoville}, {Murchikova}, {Walter},
  {Vlahakis}, {Koda}, {Vanden Bout}, {Barnes}, {Hernquist}, {Sheth}, {Yun},
  {Sanders}, {Armus}, {Cox}, {Thompson}, {Robertson}, {Zschaechner}, {Tacconi},
  {Torrey}, {Hayward}, {Genzel}, {Hopkins}, {van der Werf}, \&
  {Decarli}}]{scoville_2017_arp220}
{Scoville}, N., {Murchikova}, L., {Walter}, F., {et~al.} 2017, \apj, 836, 66

\bibitem[{{Scoville} {et~al.}(2016){Scoville}, {Sheth}, {Aussel}, {Vanden
  Bout}, {Capak}, {Bongiorno}, {Casey}, {Murchikova}, {Koda},
  {{\'A}lvarez-M{\'a}rquez}, {Lee}, {Laigle}, {McCracken}, {Ilbert}, {Pope},
  {Sanders}, {Chu}, {Toft}, {Ivison}, \& {Manohar}}]{scoville_2016}
{Scoville}, N., {Sheth}, K., {Aussel}, H., {et~al.} 2016, \apj, 820, 83

\bibitem[{{Shangguan} {et~al.}(2020){Shangguan}, {Ho}, {Bauer}, {Wang}, \&
  {Treister}}]{shangguan_2020}
{Shangguan}, J., {Ho}, L.~C., {Bauer}, F.~E., {Wang}, R., \& {Treister}, E.
  2020, \apj, 899, 112

\bibitem[{{Shanks} {et~al.}(2021){Shanks}, {Ansarinejad}, {Bielby}, {Heywood},
  {Metcalfe}, \& {Wang}}]{shanks_2021}
{Shanks}, T., {Ansarinejad}, B., {Bielby}, R.~M., {et~al.} 2021, \mnras, 505,
  1509

\bibitem[{{Sharon} {et~al.}(2016){Sharon}, {Riechers}, {Hodge}, {Carilli},
  {Walter}, {Wei{\ss}}, {Knudsen}, \& {Wagg}}]{sharon_2016}
{Sharon}, C.~E., {Riechers}, D.~A., {Hodge}, J., {et~al.} 2016, \apj, 827, 18

\bibitem[{{Silk}(2013)}]{silk_2013}
{Silk}, J. 2013, \apj, 772, 112

\bibitem[{{Simpson} {et~al.}(2017){Simpson}, {Smail}, {Wang}, {Riechers},
  {Dunlop}, {Ao}, {Bourne}, {Bunker}, {Chapman}, {Chen}, {Dannerbauer},
  {Geach}, {Goto}, {Harrison}, {Hwang}, {Ivison}, {Kodama}, {Lee}, {Lee},
  {Lee}, {Lim}, {Micha{\l}owski}, {Rosario}, {Shim}, {Shu}, {Swinbank}, {Tee},
  {Toba}, {Valiante}, {Wang}, \& {Zheng}}]{simpson_2017}
{Simpson}, J.~M., {Smail}, I., {Wang}, W.-H., {et~al.} 2017, \apjl, 844, L10

\bibitem[{{Solomon} \& {Vanden Bout}(2005)}]{solomon_2005}
{Solomon}, P.~M. \& {Vanden Bout}, P.~A. 2005, \araa, 43, 677

\bibitem[{{Spilker} {et~al.}(2014){Spilker}, {Marrone}, {Aguirre}, {Aravena},
  {Ashby}, {B{\'e}thermin}, {Bradford}, {Bothwell}, {Brodwin}, {Carlstrom},
  {Chapman}, {Crawford}, {de Breuck}, {Fassnacht}, {Gonzalez}, {Greve},
  {Gullberg}, {Hezaveh}, {Holzapfel}, {Husband}, {Ma}, {Malkan}, {Murphy},
  {Reichardt}, {Rotermund}, {Stalder}, {Stark}, {Strandet}, {Vieira},
  {Wei{\ss}}, \& {Welikala}}]{spilker_2014}
{Spilker}, J.~S., {Marrone}, D.~P., {Aguirre}, J.~E., {et~al.} 2014, \apj, 785,
  149

\bibitem[{{Spilker} {et~al.}(2016){Spilker}, {Marrone}, {Aravena},
  {B{\'e}thermin}, {Bothwell}, {Carlstrom}, {Chapman}, {Crawford}, {de Breuck},
  {Fassnacht}, {Gonzalez}, {Greve}, {Hezaveh}, {Litke}, {Ma}, {Malkan},
  {Rotermund}, {Strandet}, {Vieira}, {Weiss}, \& {Welikala}}]{spilker_2016}
{Spilker}, J.~S., {Marrone}, D.~P., {Aravena}, M., {et~al.} 2016, \apj, 826,
  112

\bibitem[{{Stanley} {et~al.}(2017){Stanley}, {Alexander}, {Harrison},
  {Rosario}, {Wang}, {Aird}, {Bourne}, {Dunne}, {Dye}, {Eales}, {Knudsen},
  {Micha{\l}owski}, {Valiante}, {De Zotti}, {Furlanetto}, {Ivison}, {Maddox},
  \& {Smith}}]{stanley_2017}
{Stanley}, F., {Alexander}, D.~M., {Harrison}, C.~M., {et~al.} 2017, \mnras,
  472, 2221

\bibitem[{{Stanley} {et~al.}(2015){Stanley}, {Harrison}, {Alexander},
  {Swinbank}, {Aird}, {Del Moro}, {Hickox}, \& {Mullaney}}]{stanley_2015}
{Stanley}, F., {Harrison}, C.~M., {Alexander}, D.~M., {et~al.} 2015, \mnras,
  453, 591

\bibitem[{{Symeonidis}(2017)}]{symeonidis_2017}
{Symeonidis}, M. 2017, \mnras, 465, 1401

\bibitem[{{Symeonidis} {et~al.}(2016){Symeonidis}, {Giblin}, {Page}, {Pearson},
  {Bendo}, {Seymour}, \& {Oliver}}]{symeonidis_2016}
{Symeonidis}, M., {Giblin}, B.~M., {Page}, M.~J., {et~al.} 2016, \mnras, 459,
  257

\bibitem[{{Tacconi} {et~al.}(2018){Tacconi}, {Genzel}, {Saintonge}, {Combes},
  {Garc{\'{\i}}a-Burillo}, {Neri}, {Bolatto}, {Contini}, {F{\"o}rster
  Schreiber}, {Lilly}, {Lutz}, {Wuyts}, {Accurso}, {Boissier}, {Boone},
  {Bouch{\'e}}, {Bournaud}, {Burkert}, {Carollo}, {Cooper}, {Cox}, {Feruglio},
  {Freundlich}, {Herrera-Camus}, {Juneau}, {Lippa}, {Naab}, {Renzini},
  {Salome}, {Sternberg}, {Tadaki}, {{\"U}bler}, {Walter}, {Weiner}, \&
  {Weiss}}]{tacconi_2018}
{Tacconi}, L.~J., {Genzel}, R., {Saintonge}, A., {et~al.} 2018, \apj, 853, 179

\bibitem[{{Talia} {et~al.}(2018){Talia}, {Pozzi}, {Vallini}, {Cimatti},
  {Cassata}, {Fraternali}, {Brusa}, {Daddi}, {Delvecchio}, \&
  {Ibar}}]{talia_2018}
{Talia}, M., {Pozzi}, F., {Vallini}, L., {et~al.} 2018, \mnras, 476, 3956

\bibitem[{{Torrey} {et~al.}(2020){Torrey}, {Hopkins}, {Faucher-Gigu{\`e}re},
  {Angl{\'e}s-Alc{\'a}zar}, {Quataert}, {Ma}, {Feldmann}, {Keres}, \&
  {Murray}}]{torrey_2020}
{Torrey}, P., {Hopkins}, P.~F., {Faucher-Gigu{\`e}re}, C.-A., {et~al.} 2020,
  \mnras, 497, 5292

\bibitem[{{Valentino} {et~al.}(2020{\natexlab{a}}){Valentino}, {Daddi},
  {Puglisi}, {Magdis}, {Liu}, {Kokorev}, {Cortzen}, {Madden}, {Aravena},
  {G{\'o}mez-Guijarro}, {Lee}, {Le Floc'h}, {Gao}, {Gobat}, {Bournaud},
  {Dannerbauer}, {Jin}, {Dickinson}, {Kartaltepe}, \&
  {Sanders}}]{valentino_2020c}
{Valentino}, F., {Daddi}, E., {Puglisi}, A., {et~al.} 2020{\natexlab{a}}, \aap,
  641, A155

\bibitem[{{Valentino} {et~al.}(2018){Valentino}, {Magdis}, {Daddi}, {Liu},
  {Aravena}, {Bournaud}, {Cibinel}, {Cormier}, {Dickinson}, {Gao}, {Jin},
  {Juneau}, {Kartaltepe}, {Lee}, {Madden}, {Puglisi}, {Sanders}, \&
  {Silverman}}]{valentino_2018}
{Valentino}, F., {Magdis}, G.~E., {Daddi}, E., {et~al.} 2018, \apj, 869, 27

\bibitem[{{Valentino} {et~al.}(2020{\natexlab{b}}){Valentino}, {Magdis},
  {Daddi}, {Liu}, {Aravena}, {Bournaud}, {Cortzen}, {Gao}, {Jin}, {Juneau},
  {Kartaltepe}, {Kokorev}, {Lee}, {Madden}, {Narayanan}, {Popping}, \&
  {Puglisi}}]{valentino_2020b}
{Valentino}, F., {Magdis}, G.~E., {Daddi}, E., {et~al.} 2020{\natexlab{b}},
  \apj, 890, 24

\bibitem[{{Vallini} {et~al.}(2019){Vallini}, {Tielens}, {Pallottini},
  {Gallerani}, {Gruppioni}, {Carniani}, {Pozzi}, \& {Talia}}]{vallini_2019}
{Vallini}, L., {Tielens}, A.~G.~G.~M., {Pallottini}, A., {et~al.} 2019, \mnras,
  490, 4502

\bibitem[{{van der Wel} {et~al.}(2014){van der Wel}, {Franx}, {van Dokkum},
  {Skelton}, {Momcheva}, {Whitaker}, {Brammer}, {Bell}, {Rix}, {Wuyts},
  {Ferguson}, {Holden}, {Barro}, {Koekemoer}, {Chang}, {McGrath},
  {H{\"a}ussler}, {Dekel}, {Behroozi}, {Fumagalli}, {Leja}, {Lundgren},
  {Maseda}, {Nelson}, {Wake}, {Patel}, {Labb{\'e}}, {Faber}, {Grogin}, \&
  {Kocevski}}]{vanderwel_2014}
{van der Wel}, A., {Franx}, M., {van Dokkum}, P.~G., {et~al.} 2014, \apj, 788,
  28

\bibitem[{{van der Werf} {et~al.}(2010){van der Werf}, {Isaak}, {Meijerink},
  {Spaans}, {Rykala}, {Fulton}, {Loenen}, {Walter}, {Wei{\ss}}, {Armus},
  {Fischer}, {Israel}, {Harris}, {Veilleux}, {Henkel}, {Savini}, {Lord},
  {Smith}, {Gonz{\'a}lez-Alfonso}, {Naylor}, {Aalto}, {Charmandaris}, {Dasyra},
  {Evans}, {Gao}, {Greve}, {G{\"u}sten}, {Kramer}, {Mart{\'\i}n-Pintado},
  {Mazzarella}, {Papadopoulos}, {Sanders}, {Spinoglio}, {Stacey}, {Vlahakis},
  {Wiedner}, \& {Xilouris}}]{vanderwerf_2010}
{van der Werf}, P.~P., {Isaak}, K.~G., {Meijerink}, R., {et~al.} 2010, \aap,
  518, L42

\bibitem[{{Vasudevan} \& {Fabian}(2007)}]{vasudevan_2007}
{Vasudevan}, R.~V. \& {Fabian}, A.~C. 2007, \mnras, 381, 1235

\bibitem[{{Veilleux} {et~al.}(2020){Veilleux}, {Maiolino}, {Bolatto}, \&
  {Aalto}}]{veilleux_2020}
{Veilleux}, S., {Maiolino}, R., {Bolatto}, A.~D., \& {Aalto}, S. 2020, \aapr,
  28, 2

\bibitem[{{Walter} {et~al.}(2011){Walter}, {Wei{\ss}}, {Downes}, {Decarli}, \&
  {Henkel}}]{walter_2011}
{Walter}, F., {Wei{\ss}}, A., {Downes}, D., {Decarli}, R., \& {Henkel}, C.
  2011, \apj, 730, 18

\bibitem[{{Wei{\ss}} {et~al.}(2005){Wei{\ss}}, {Downes}, {Henkel}, \&
  {Walter}}]{weiss_2005}
{Wei{\ss}}, A., {Downes}, D., {Henkel}, C., \& {Walter}, F. 2005, \aap, 429,
  L25

\bibitem[{{Wei{\ss}} {et~al.}(2007){Wei{\ss}}, {Downes}, {Neri}, {Walter},
  {Henkel}, {Wilner}, {Wagg}, \& {Wiklind}}]{weiss_2007}
{Wei{\ss}}, A., {Downes}, D., {Neri}, R., {et~al.} 2007, \aap, 467, 955

\bibitem[{{Weiss} {et~al.}(2007){Weiss}, {Downes}, {Walter}, \&
  {Henkel}}]{weiss_2007proceedings}
{Weiss}, A., {Downes}, D., {Walter}, F., \& {Henkel}, C. 2007, in Astronomical
  Society of the Pacific Conference Series, Vol. 375, From Z-Machines to ALMA:
  (Sub)Millimeter Spectroscopy of Galaxies, ed. A.~J. {Baker}, J.~{Glenn},
  A.~I. {Harris}, J.~G. {Mangum}, \& M.~S. {Yun}, 25

\bibitem[{{Yesuf} \& {Ho}(2020)}]{yesuf_2020}
{Yesuf}, H.~M. \& {Ho}, L.~C. 2020, \apj, 901, 42

\end{thebibliography}
\bibliographystyle{aa}

\appendix
\section{Data table}
\label{app:table}
The data used in this work are release in electronic format. The
description of the table columns is reported in Table
\ref{tab:datatable}. More information can be found in the master table
released in \cite{valentino_2020c} (e.g., ALMA continuum emission
measurements). Whenever a difference between this and the master table
is present, the updated estimates in this work supersede the previous
version. Significant variations exclusively concern the far-IR SED
modeling of a handful of AGN with extreme \fagn, as described in
Sect. \ref{sec:data}. 

\begin{table*}
\small
  \centering
  \caption{Column description for the data release.}
  \begin{tabular}{lcl}
    \toprule
    \toprule
    Name & Units & Description\\
    \midrule
    ID& \dots& Identifier \citep{valentino_2020c}\\
    R.A.& hh:mm:ss& Right ascension\\
    Dec&  dd:mm:ss& Declination\\
    zspec\textunderscore opticalnir& \dots& Optical or near-IR spectroscopic redshift (M. Salvato et al., in prep.)\\ 
    (d)zspec\textunderscore submm& \dots& ALMA submillimeter spectroscopic redshift\\ 
    Log\textunderscore StellarMass& \msun& Logarithm of the stellar mass (\citealt{chabrier_2003} IMF, uncertainty: 0.2 dex)\\
    (d)Total\textunderscore LIR& \lsun& Total $8-1000$ $\mu$m \lir\ \citep{draine_2007, mullaney_2011}\\
    (d)SF\textunderscore LIR& \lsun& $L_{\rm IR,\,SFR}$ from the star-forming component ($L_{\rm IR,\,SFR}=L_{\rm IR}-L_{\rm IR,\,AGN}$)\\
    (d)AGN\textunderscore LIR& \lsun& $L_{\rm IR,\,AGN}$ from the AGN component ($L_{\rm IR,\,AGN}=L_{\rm IR}\times f_{\rm AGN})$\\
    (d)$f_{\rm AGN}$& \dots& Fraction of \lir\ due to the AGN emission\\
    (d)$M_{\rm dust}$& \msun& Dust mass \citep{draine_2007}\\
    (d)$U$& \dots& Mean intensity of the interstellar radiation field \citep{draine_2007}\\
    DistanceMS & \dots& Distance from the main sequence as parameterized in \cite{sargent_2014}\\
    (d)Size& arcsec& Source angular size from ALMA\\
    OneSigma\textunderscore Size& arcsec& $1\sigma$ upper limit on the source angular size from ALMA\\ 
    Flux\textunderscore Line($X$)& \jykms& Velocity integrated flux of line $X$\\
    SNR\textunderscore ($X$)& \dots& Signal to noise ratio of the flux of line $X$\\
    OneSigma\textunderscore ($X$)& \jykms& $1\sigma$ upper limit on the flux of line $X$\\
    Width\textunderscore ($X$)& \kms& Velocity width of line $X$\\
    Flag\textunderscore ($X$)& \dots& Quality and usage flag for line $X$ (Flag=1: detection; Flag=0.5: robust upper limit)\\
    Lx\textunderscore 210& \es& 2-10 keV rest frame logarithmic luminosity (column \#50 in \citealt{marchesi_2016}; negative values: upper limits)\\
    Flag\textunderscore Donley2012& \dots& \cite{donley_2012} AGN classification (Flag=1: AGN; Flag=0: non-AGN)\\

    \midrule
    \bottomrule
  \end{tabular}
  \tablefoot{\smallskip This full table is available at the CDS. More information can be found in \cite{valentino_2020c}. Whenever a difference is present, the updated estimates in this table supersede the previous version. 
  
  - Lines $X$: \cofive, \cotwo, \coseven, \citwo, \cione, and \cofour.\\
  - The uncertainties have the same name of the quantity that they refer to,
  preceded by $d$ (e.g., Total\textunderscore LIR $\pm$ dTotal\textunderscore LIR). Values $-99$ indicate the absence of data for a specific quantity.\\
  - The \cofive\ and \cotwo\ fluxes, \mstar, and the angular size estimates for ID=51280 are from \cite{brusa_2018}.}
  \label{tab:datatable}
\end{table*}

\section{Supplementary figures}
\label{app:supp_figures}
We collect here further plots that might help guide the reader
  through more technical aspects of the analysis in this work.

\subsection{Correlation between the X-ray luminosity and IR emission
  in AGN hosts}
\begin{figure}
  \centering
  \includegraphics[width=\columnwidth]{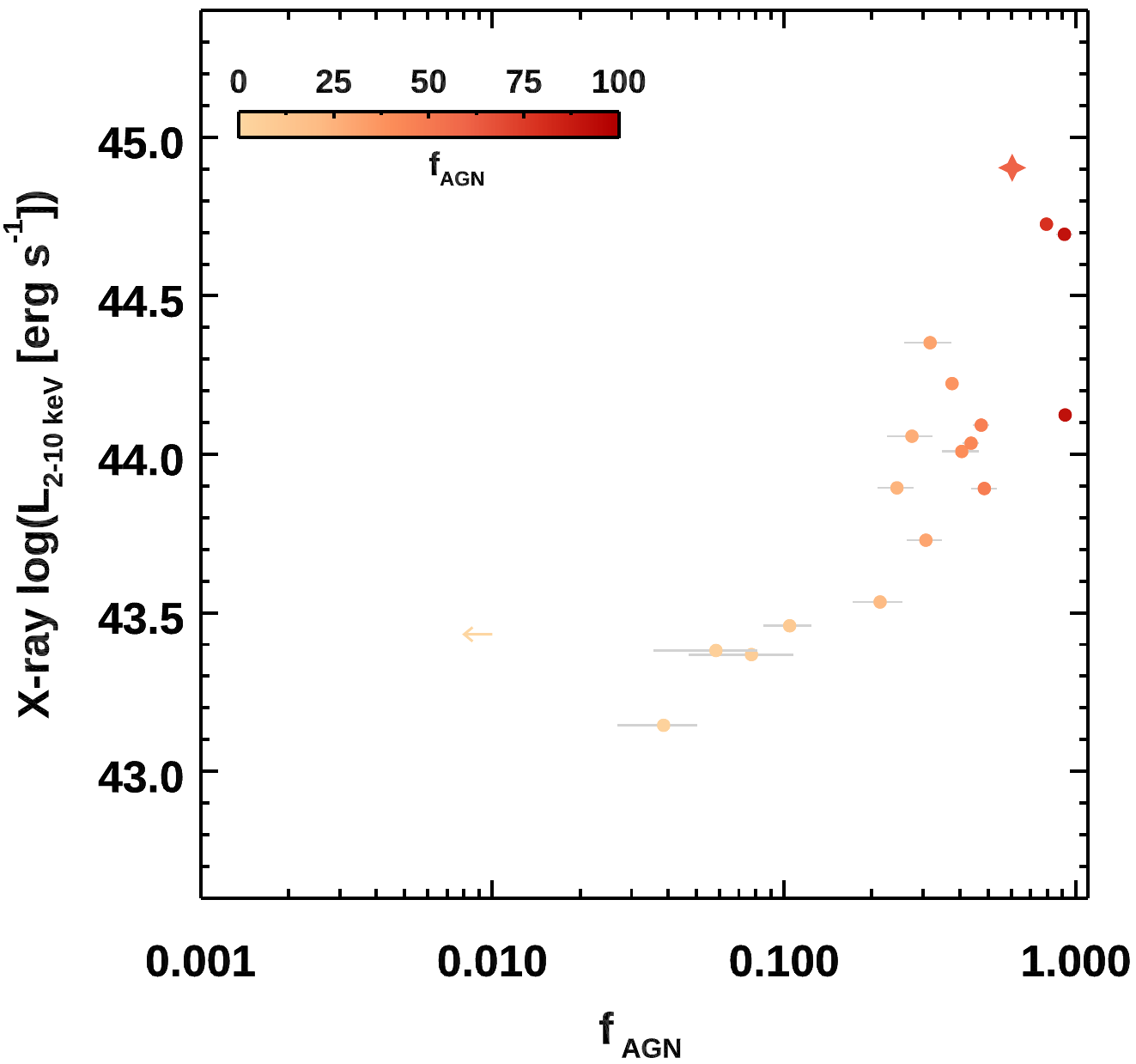}
  \caption{{X-ray luminosity \lx\ as a function of the AGN
      contribution to the far-IR emission.} The symbols represent our
    sample at $z\sim1.2$ and are color coded according to \fagn\ as in
    previous figures. The arrow marks the floor of $f_{\rm AGN}=1$\%
    adopted in this work. The solid orange star indicates the object from
  \cite{brusa_2018}, which is part of our parent sample.}
  \label{fig:app:fagnxray}
  \end{figure}
In Fig. \ref{fig:app:fagnxray} we show the relation
  between the AGN contribution to the total IR luminosity $f_{\rm
    AGN}=L_{\rm IR,\,AGN}/L_{\rm IR}$ and the X-ray luminosity in the 2-10
keV band for the sources in our sample detected by \textit{Chandra}
\citep{civano_2016, marchesi_2016}. The correlation supports the use of both
variables as indicators of AGN activity and justifies our choice to
fix the \fagn+$1\sigma_{f_{\rm AGN}}$ threshold at 20\% to broadly classify objects as AGN
hosts or star-forming galaxies (Sect. \ref{sec:effectsed}).

\subsection{Stellar masses}
\begin{figure*}
  \includegraphics[width=\textwidth]{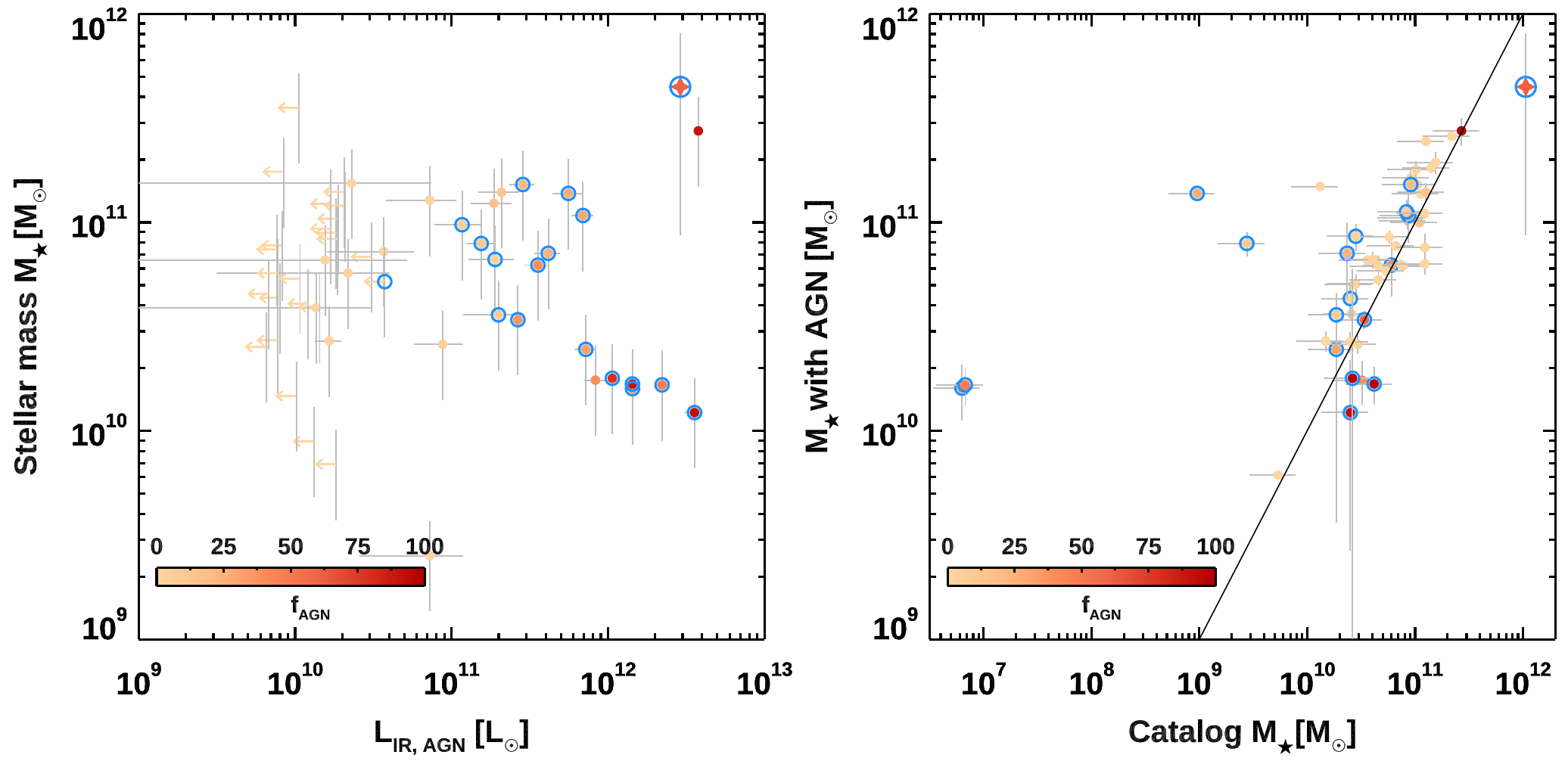}
  \caption{{Detectability of AGN signatures at different stellar
      masses.} \textit{Left:} Stellar mass as a function of the IR luminosity
    attributed to the AGN emission \liragn. The symbols represent our
    sample at $z\sim1.2$ and are color coded according to \fagn\ as in
    previous figures. Empty blue circles indicate X-ray
    detections. Arrows mark the floor of $f_{\rm AGN}=1$\% adopted in
    this work. A similar picture is offered when using \lx\ as a proxy
    for the AGN emission in lieu of \liragn. \textit{Right:} Stellar
    masses from the standard stellar population synthesis modeling
    \citep{muzzin_2013, laigle_2016} compared with results obtained by
    following \cite{circosta_2018}. Symbols and colors are the same as in the left panel.} 
  \label{fig:app:mass}
\end{figure*}
In Fig. \ref{fig:app:mass}, we show the stellar masses for our
sample of $z\sim1.2$ galaxies as a function of \liragn. Despite the
large scatter, AGN hosts appear to have lower stellar masses in our
sample, reflecting the IR selection and AGN classification (Sect.
\ref{sec:effectsed}). We stress that deriving stellar masses for
optically bright AGN is prone to significant uncertainties. In Fig.
\ref{fig:app:mass}, we also show the comparison between estimates of
\mstar\ from the public COSMOS catalogs \citep{muzzin_2013,
  laigle_2016} against a refit with \textsc{Cigale} that includes AGN
templates as in \cite{circosta_2018}. For the object presented in
  \cite{brusa_2018}, we used their original fit including AGN templates
  (\citealt{perna_2015} and references therein). The AGN component is
allowed for every source with AGN signatures from X-ray or mid-IR
 observations. While not significantly introducing
any systematic bias for star-forming galaxies, the use of a code that
allows for a consistent AGN treatment improves the stellar mass
estimates of their hosts. This happens both removing the strongest and unphysical
outliers at low \mstar\ and reducing
the estimates for the objects with the largest \fagn\ values due to the
proper decomposition of the AGN and stellar contributions. 

\subsection{Individual spectral line energy distributions}
\begin{figure}
\includegraphics[width=\columnwidth]{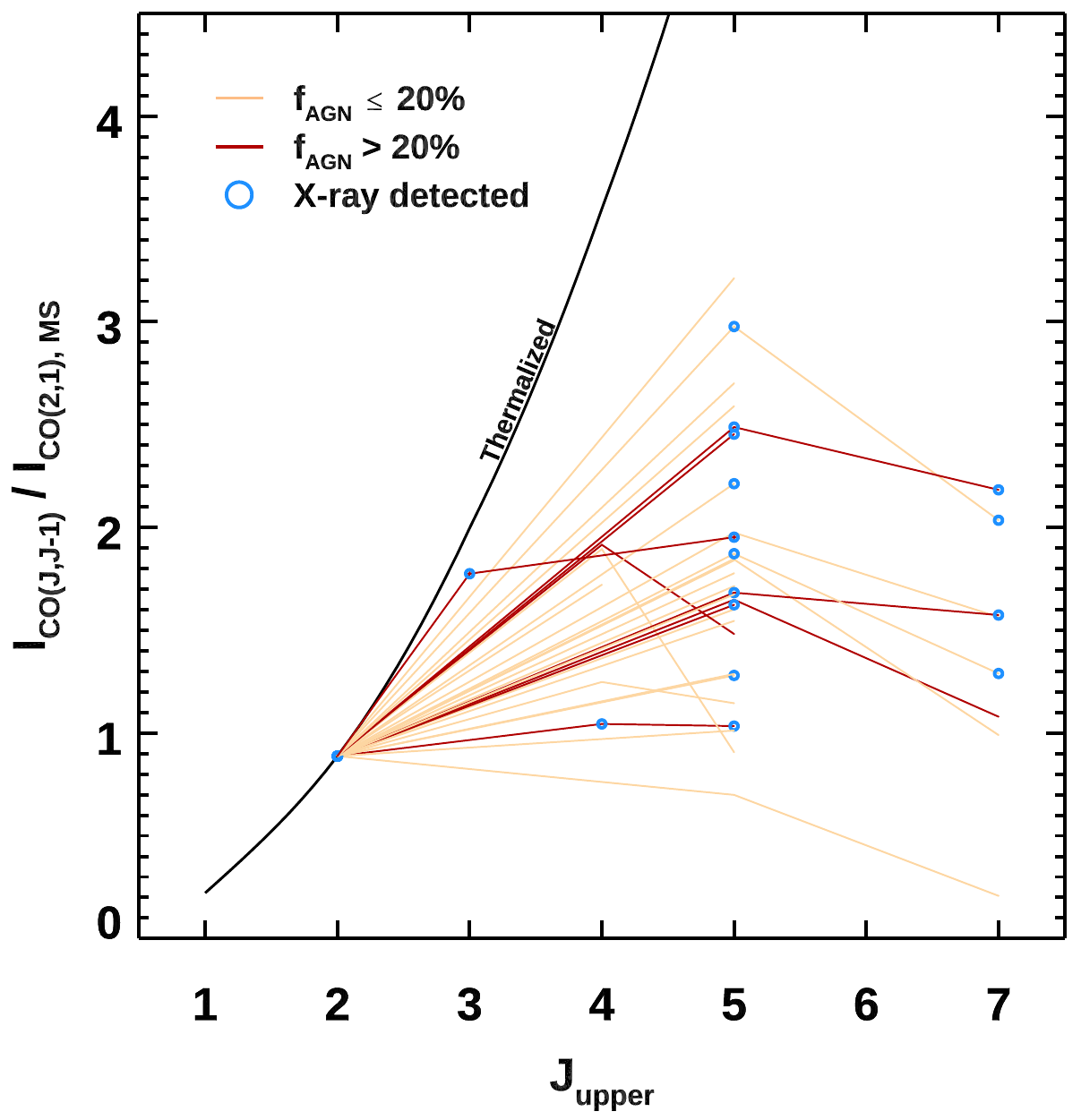}
\caption{Individual CO spectral line distributions of AGN hosts and
    star-forming galaxies and individual SLEDs for AGN hosts and star-forming
  galaxies normalized to the average \cotwo\ flux of main-sequence
  galaxies at 
  $z\sim1.2$ from V20a. Yellow and red
  lines indicate galaxies in our sample with $f_{\rm
      AGN}\leq20$\% and $>20$\%, respectively. Open blue
  circles show \textit{Chandra} detections at 2-10 keV.}
\label{fig:individualsleds}
\end{figure}

In Fig. \ref{fig:individualsleds} we show the individual CO
  SLEDs of all the \cotwo\ detected objects in this study, classified
  as AGN hosts and star-forming galaxies according to a threshold in
  $f_{\rm AGN}$ of 20\% and a \textit{Chandra} detection at 2-10 keV.

\begin{table*}
  \caption{Average SLEDs for star-forming
    galaxies and AGN hosts at $z\sim1.25$.}
  \begin{tabular}{ccccccccccccc}
    \toprule
    \toprule
     & \multicolumn{3}{c}{X-ray} & \multicolumn{3}{c}{No X-ray}\\
    \cmidrule(lr){2-4}  
    \cmidrule(lr){5-7}
    Transition & $N_{\rm det}, N_{\rm up}$& Mean& Median &
                         $N_{\rm det}, N_{\rm up}$& Mean& Median \\
    \midrule
    \lprimecotwo&    $10, 4$&  \tablefootmark{$\dagger$}$1.46^{+0.16}_{-0.16}$& $1.49^{+0.38}_{-0.89}$&
                                $21, 4$&  \tablefootmark{$\dagger$}$1.87^{+0.23}_{-0.23}$& $1.65^{+0.65}_{-0.82}$\\

    \lprimecofour&   $1, 0$  &  $-$& $-$&
                                $4, 0$  &  $0.63^{+0.11}_{-0.11}$& $0.71^{+0.16}_{-0.12}$\\

    \lprimecofive&    $15, 1$&  \tablefootmark{$\dagger$}$0.48^{+0.07}_{-0.07}$& $0.42^{+0.16}_{-0.15}$&
                                $26, 1$  &  $0.56^{+0.06}_{-0.06}$& $0.45^{+0.39}_{-0.10}$\\

    \lprimecoseven& $6, 0$  &  $0.23^{+0.04}_{-0.04}$& $0.22^{+0.06}_{-0.04}$ &
                                $6, 0$  &  $0.16^{+0.04}_{-0.04}$& $0.15^{+0.04}_{-0.05}$\\

    \midrule
    \lprimecione&     $2, 0$  &  $0.36^{+0.12}_{-0.12}$& $-$ &
                                $8, 1$  &  $0.34^{+0.05}_{-0.05}$& $0.29^{+0.11}_{-0.10}$\\

    \lprimecitwo&     $6, 0$  &   $0.21^{+0.04}_{-0.04}$& $0.19^{+0.05}_{-0.04}$&
                                $6, 0$  &  $0.16^{+0.03}_{-0.03}$& $0.18^{+0.06}_{-0.07}$\\

    \midrule\\
     & \multicolumn{3}{c}{$f_{\rm AGN} > 20$\%} & \multicolumn{3}{c}{$f_{\rm AGN} \leq 20$\%} \\
    \cmidrule(lr){2-4}  
    \cmidrule(lr){5-7}
    Transition & $N_{\rm det}, N_{\rm up}$& Mean& Median &
                         $N_{\rm det}, N_{\rm up}$& Mean& Median \\
    \midrule
     \lprimecotwo&   $8, 4$&  \tablefootmark{$\dagger$}$1.23^{+0.16}_{-0.16}$& $0.94^{+0.69}_{-0.29}$&
                                $23, 4$&  \tablefootmark{$\dagger$}$1.94^{+0.21}_{-0.21}$& $1.68^{+0.69}_{-0.80}$\\
     \lprimecofour&   $2, 0$&  $0.41^{+0.06}_{-0.06}$& $-$&
                                $3, 0$&  $0.73^{+0.08}_{-0.08}$& $0.71^{+0.16}_{-0.12}$\\
     \lprimecofive&   $13, 0$&  $0.40^{+0.04}_{-0.04}$& $0.38^{+0.14}_{-0.15}$&
                                $28, 2$&  \tablefootmark{$\dagger$}$0.58^{+0.06}_{-0.06}$& $0.45^{+0.40}_{-0.10}$\\
     \lprimecoseven&   $4, 0$&  $0.21^{+0.02}_{-0.02}$& $0.19^{+0.09}_{-0.01}$&
                                   $8, 0$&  $0.19^{+0.05}_{-0.05}$& $0.15^{+0.19}_{-0.04}$\\
    \midrule
    \lprimecione&     $3, 0$  &  $0.30^{+0.09}_{-0.09}$& $0.24^{+0.25}_{-0.07}$&
                                $7, 1$  &  \tablefootmark{$\dagger$}$0.37^{+0.05}_{-0.05}$& $0.31^{+0.09}_{-0.08}$\\
    \lprimecitwo&     $4, 0$&  $0.19^{+0.02}_{-0.02}$& $0.19^{+0.05}_{-0.02}$&
                                $8, 0$  &  $0.18^{+0.04}_{-0.04}$& $0.19^{+0.05}_{-0.09}$\\
    \bottomrule
  \end{tabular}
  \tablefoot{The $L'$ luminosities are expressed in $10^{10}$ \kkmspc. The
  average $I$ fluxes in \jykms\ shown in Fig. \ref{fig:fig4} are computed adopting $z=1.25$. The uncertainty on the median value is the interquartile range.}
  \tablefoottext{$\dagger$}{Formally biased mean value as the first upper
  limit was turned into a detection for the calculation of the KM
  estimator \citep{kaplan-meier_1958}.}
  \label{tab:sled}
\end{table*}

\end{document}